\documentstyle[psfig,11pt]{article}

\input amssym.def
\input amssym.tex

\def\bbbone{{\mathchoice {\rm 1\mskip-4mu l} {\rm 1\mskip-4mu l}    
{\rm 1\mskip-4.5mu l} {\rm 1\mskip-5mu l}}}

\def\NN{\mbox{\rm I \hspace{-0.87em} N}}
\def\ZZ{\mbox{\sf Z \hspace{-1.12em} Z}}

\def\RR{\mbox{\rm I \hspace{-0.87em} R}}
\def\CC{\mbox{{\sf I} \hspace{-1.13em} C}}

\title{Mean Green's function of the Anderson model \\ 
at weak disorder with an infra-red cut-off}
\author{Gilles Poirot\\
Centre de Physique Th{\'e}orique, Ecole Polytechnique\\
91128 Palaiseau Cedex, FRANCE\\
poirot@cpth.polytechnique.fr}

\begin{document}
\sloppy
\maketitle
\begin{abstract}
In this paper we develop a polymer expansion with large/small field conditions
for the mean resolvent of a weakly disordered system. Then we show that we can
apply our result to a two-dimensional model, for energies outside 
the unperturbed spectrum or in the free spectrum provided the potential 
has an infra-red cut-off. This leads to an asymptotic expansion for the 
density of states. 
We believe this is an important first step towards a rigorous
analysis of the density of states in the free spectrum of a random Schr{\"o}dinger
operator at weak disorder.  
\end{abstract}

\newtheorem{theorem}{Theorem}
\newtheorem{lemma}{Lemma}


\section{Introduction}
In the one-body approximation, the study of disordered systems
amounts to the study of random Schr{\"o}dinger operators of the form
\begin{equation}
  H = H_{0} + \lambda V
\end{equation}
where $H_{0}$ is a kinetic term ({\it i.e.} a self-adjoint or essentially
self-adjoint operator corresponding to
some dispersion relation, typically a regularized version of $-\Delta$) 
and $V$ is a real random potential (in the
simplest case, $V$ is a white noise). We work on a ultra-violet regular
subspace of ${\cal L}^{2} (\mbox{I}\!\mbox{R}^{d})$ 
and we restrict ourselves to $\lambda$ small 
so as to see $\lambda V$ as a kind of perturbation of the free Hamiltonian. 

The properties of $H$ are usually established through the behavior of the
kernel of the resolvent operator or Green's function (\cite{Tho}, \cite{Aiz},
\cite {Fra})
\begin{equation}
  G_{E}(x, y) = \, < \! x | \frac{1}{H-E} | y \! >
\end{equation}
For instance, the density of states is given by
\begin{equation}
  \rho(E) = \frac{1}{\pi} \lim_{\varepsilon \rightarrow 0} \mbox{Im } 
    G_{E+i \varepsilon}(x, x)
\end{equation} 

The important point is that, in the thermodynamic limit, the system is
self-averaging, {\em i.e.} mean properties are often almost sure ones. Thus
the problem can be seen as a statistical field theory with respect to the
random field $V$. In Statistical Mechanics, functional integrals in the 
weakly coupled regime are controled through a cluster expansion 
(or polymer expansion) with small field versus large field conditions, the
problem being then to control a Boltzmann weight (\cite{Bry}, \cite{Riv1}). 

In the first part of this paper, we derive a resolvent cluster expansion with
large field versus small field conditions assuming that $V$ satisfies some 
large deviation estimates. This would allow to prove the existence and the 
regularity of the mean Green's function (theorem \ref{thRE}) and to get an 
asymptotic expansion for the density of states. 

In the second part, we show that the hypothesis of theorem \ref{thRE} are
satisfied in the case of a 2 dimensional model with a rotation invariant
dispersion relation and an infra-red cut-off on the potential. From the point
of view of {\em Renormalization Group} analysis, our results
allows to control the model away from the singularity, {\em i.e.} to  perform
the first renormalization group steps and therefore to generate a fraction of
the expected ``mass''.


\section{Model and results}
\subsection{The model}
In $\mbox{I}\!\mbox{R}^{d}$ we consider 
\begin{equation}
  H = H_{0} + \lambda V  
\end{equation}
where $V$ is a gaussian random field with covariance $\xi$ whose 
smooth translation invariant kernel is rapidly decaying 
(we will note the associated measure $d\mu_{\xi}$). Because
$\xi$ is smooth, $d\mu_{\xi}$ as a measure on tempered distributions is in
fact supported on ${\cal C}^{\infty}$ functions.
We suppose also that $\hat{H}_{0}^{-1}$ has compact support so that we do not
have to deal with ultra-violet problems.  
We construct the finite volume model in $\mbox{I}\!\mbox{R}^{d}/_{\! 
\displaystyle \Lambda \mbox{\sf Z\hspace{-5pt}Z}^{\scriptstyle d}}$ 
by replacing $\xi$ and $H_{0}$ by their ``$\Lambda$-periodization''  
\begin{eqnarray}
  \xi_{\Lambda}(x,y) &=& \frac{1}{\Lambda^{d}} \sum_{p \in
    \frac{2\pi}{\Lambda}\mbox{\scriptsize \sf Z\hspace{-3.5pt}Z}^{d}} 
    e^{ip(x-y)} \hat{\xi}(p) = \sum_{z \in \Lambda \mbox{\scriptsize \sf
    Z\hspace{-3.5pt}Z}^{d}} \xi(x-y+z) \\
  H_{0}^{(\Lambda)}(x,y) &=& \ldots = \sum_{z \in \Lambda \mbox{\scriptsize \sf
    Z\hspace{-3.5pt}Z}^{d}} H_{0}(x-y+z) 
\end{eqnarray}

Then we define
\begin{eqnarray}
  G_{\Lambda, \varepsilon} (E, \lambda, V) &=& 
    \frac{1}{H_{0}^{(\Lambda)} + \lambda V - (E + i\varepsilon)} \\ 
  G_{\Lambda, \varepsilon} (E, \lambda) &=& \int \! d\mu_{\xi_{\Lambda}}(V) \, 
    G_{\Lambda, \varepsilon} (E, \lambda, V)
\end{eqnarray}  
where $d\mu_{\xi_{\Lambda}}$ can be considered either as a
measure on ${\cal C}^{\infty}\! \left(\mbox{I}\!\mbox{R}^{d}/_{\! 
\displaystyle \Lambda \mbox{\sf Z\hspace{-5pt}Z}^{\scriptstyle d}} \right)$ or
as a measure on ${\cal C}^{\infty} (\mbox{I}\!\mbox{R}^{d})$ which is
supported by the space of $\Lambda$-periodic functions. In the same way,
$G_{\Lambda, \varepsilon}$ will be considered as an operator either on 
${\cal L}^{2} \! \left(\mbox{I}\!\mbox{R}^{d}/_{\! \displaystyle \Lambda 
\mbox{\sf Z\hspace{-5pt}Z}^{\scriptstyle d}} \right)$ or on ${\cal
L}^{2}(\mbox{I}\!\mbox{R}^{d})$. One can note that in momentum
space, because of the cut-off, the problem reduces to a finite dimensional
one. 

Because $V$ is almost surely regular, its operator norm as a multiplicative 
operator is equal to its ${\cal L}^{\infty}$ norm (it is easy to see that $\|
V\| \leqslant \| V\|_{\infty}$, equality can be obtained by taking test
functions $f_{n, x}$ such that $f_{n, x}^{2} \rightarrow \delta_{x}$). 
Therefore $V$ is bounded and self-adjoint. 
Then $G_{\Lambda, \varepsilon} (E, \lambda, V)$ is almost
surely an analytic operator-valued 
function of $\lambda$ in a small domain (depending on $V$) around the
origin. This domain can be extended to a $V$-dependent neighborhood of the
real axis thanks to the identity (for $|\lambda - \mu|$ small enough)
\begin{equation}
  G_{\Lambda, \varepsilon} (E, \mu, V) = G_{\Lambda, \varepsilon} (E, \lambda,
    V) \left\{ I + \sum_{n=1}^{\infty} (\lambda- \mu)^{n} \left[V G_{\Lambda,
    \varepsilon} (E, \lambda, V) \right]^{n} \right\}
\end{equation} 
In the same way, $G_{\Lambda, \varepsilon}(E, \lambda, V)$ is analytic in
$E$. One can also check 
that $G_{\Lambda, \varepsilon}(E, \lambda, V)$ has a 
smooth kernel and is integrable with respect to
$d\mu_{\xi_{\Lambda}}$. Furthermore, $G_{\Lambda, \varepsilon}(E, \lambda)$
will have a translation invariant kernel because $d\mu_{\xi_{\Lambda}}$ is
translation invariant. 


\subsection{Main result}
We introduce a function $\theta$ which satisfies 
\begin{itemize}
  \item $\theta$ is an odd ${\cal C}^{\infty}$ function, 
    increasing and bounded 
  \item for any $x$, $|\theta(x)| \leqslant |x|$
  \item for any $|x| \leqslant 1$, $\theta(x) = x$
  \item the ${\cal L}^{\infty}$ norm of its derivatives does not grow too fast 
\end{itemize} 

Then for $\mu >0$, we define the operators $C_{\Lambda, \mu}$, 
$D_{\Lambda, \mu}$ and $U_{\Lambda, \mu}$ through the Fourier transform of 
their kernel
\begin{eqnarray}
  \hat{C}_{\Lambda, \mu}^{-1}(p) &=& \hat{H}_{0}^{(\Lambda)}(p)- E -i\mu \\
  \hat{D}_{\Lambda, \mu}(p) &=& \frac{1}{\left|\theta 
    [\hat{H}_{0}^{(\Lambda)}(p)-E] - i\mu \right|^{1/2}} \\
  \hat{U}_{\Lambda, \mu}^{-1}(p) &=& \hat{D}_{\Lambda, \mu}^{2}(p)
    \hat{C}_{\Lambda, \mu}^{-1}(p) 
\end{eqnarray}

Given any characteristic length $L$ we can divide the space into cubes $\Delta$
of side $L$ and construct an associated ${\cal C}^{\infty}_{0}$ partition of
unity  
\begin{equation}
  1 = \sum_{\Delta} \chi_{\Delta} 
\end{equation}
where $\chi_{\Delta}$ has support in a close neighborhood of the cube 
$\Delta$ ({\it e.g.} on $\Delta$ and its nearest neighbors).  
This decomposition induces an orthogonal decomposition of $V$ into a sum of
fields $V_{\Delta}$ with covariance 
\begin{equation}
  \xi_{\Lambda}^{\Delta} (x,y) = \int \! dz \, \xi_{\Lambda}^{1/2}(x-z) 
    \chi_{\Delta}(z) \xi_{\Lambda}^{1/2}(z-y)
\end{equation}

For simplicity we will pretend that $\xi$ and $\xi^{1/2}$ have compact 
support, so that $V_{\Delta}$ is almost surely supported on a close
neighborhood of $\Delta$, moreover we will take that it is restricted to 
$\Delta$ and its nearest neighbors. The generalization to a fast decaying
$\xi$ can be easily obtained by decomposing each $V_{\Delta}$ over the 
various cubes and write more complicated small/large field conditions that
test the size of $V_{\Delta}$ in the various cubes. This leads to lengthy
expressions that we want to avoid. 

Finally, we note $d_{\Lambda}$ the distance in $\mbox{I}\!\mbox{R}^{d}/_{\! 
\displaystyle \Lambda \mbox{\sf Z\hspace{-5pt}Z}^{\scriptstyle d}}$
\begin{equation}
  d_{\Lambda}(x,y) = \min_{z \in \Lambda \mbox{\scriptsize \sf
    Z\hspace{-3.5pt}Z}^{d}} |x-y+z|  
\end{equation} 

In the following, $C$ or $O(1)$ will stand as generic names for constants in
order to avoid keeping track of the numerous constants that will
appear. Furthermore we will not always make the distinction between a function
and its Fourier transform but we will use $x$, $y$ and $z$ as space variables
and $p$ and $q$ as momentum variables.  

\begin{theorem}
\label{thRE} \ \\
Suppose that 
\begin{itemize} 
  \item $\xi$ is smooth and has fast decay 
  \item $\displaystyle C_{\xi} = \sup_{\Lambda} \frac{1}{2
\Lambda^{d}} \int_{[0, \Lambda]^{d}} \! \xi^{-1}_{\Lambda}(x, y) \, dx \, dy$ 
exists   
  \item for all $E \in [E_{1}, E_{2}]$ and all $\mu$, $C_{\mu}$, $D_{\mu}$ and
$U_{\mu}$ have smooth kernels with fast decay over a length scale $L$. 
  \item for all $n_{1}$, we have $C_{n_{1}}$ such that for all $\Lambda$ and
  all triplets $(\Delta_{1}, \Delta_{2}, \Delta_{3})$ 
\begin{equation} 
  \| \chi_{\Delta_{1}} D_{\Lambda, \mu} V_{\Delta_{2}} D_{\Lambda,
    \mu} \chi_{\Delta_{3}} \| \leqslant \frac{C_{n_{1}} \|D_{\Lambda, \mu}
    V_{\Delta_{2}} D_{\Lambda, \mu}\|}{\left[1 + L^{-1}
    d_{\Lambda}(\Delta_{1}, \Delta_{2})\right]^{n_{1}} \left[1 + L^{-1}
    d_{\Lambda}(\Delta_{2}, \Delta_{3})\right]^{n_{1}}}
\end{equation}
  \item there are constants $C_{0}$, $C_{1}$, $\kappa > 0$ and 
    $\alpha >0$ such that 
\begin{equation}
\label{largedev}
  \forall \Lambda \leqslant \infty, \, \forall a>1, \, 
    \forall \Delta, \quad I\!\!P_{\Lambda} \left(\| D_{\Lambda, \mu} 
    V_{\Delta} D_{\Lambda, \mu}\| \geqslant  
    a C_{0} \right) \leqslant C_{1} e^{-\kappa a^{2} L^{\alpha}} 
\end{equation}
where $I\!\!P_{\Lambda} (.)$ denote the probability with respect to the
measure $d\mu_{\xi_{\Lambda}} \equiv \otimes d\mu_{\xi_{\Lambda}^{\Delta}}$ 
($\xi_{\infty} \equiv \xi$)
\begin{equation}
  I\!\!P_{\Lambda} (X) = \int \! d\mu_{\xi_{\Lambda}}(V) \, \bbbone_{X}(V) 
    = \mu_{\xi_{\Lambda}}(X)
\end{equation} 
\end{itemize}

Then let $\mu_{0} = L^{-d/2} C_{\xi}^{1/2}$, $\mu = \lambda \mu_{0}$ and 
\begin{equation}
  T_{\Lambda, \varepsilon} = D_{\Lambda, \mu}^{-1} G_{\Lambda, \varepsilon}
    D_{\Lambda, \mu}^{-1} 
\end{equation}

For all $\lambda \leqslant \lambda_{0} = O(1)$ and for all 
$\varepsilon$ small enough (in a $\lambda$-dependent way), 
$T_{\Lambda, \varepsilon}(E, \lambda)$ is uniformly
bounded in $\Lambda$ and admits the following development (in the operator
norm sense) 
\begin{equation}
\label{Lambdadev}
  1_{\Omega_{\Lambda}} T_{\Lambda, \varepsilon}(E, \lambda) 
    1_{\Omega_{\Lambda}} =
    1_{\Omega_{\Lambda}} T(E+i\varepsilon, \lambda) 1_{\Omega_{\Lambda}} 
    + \mbox{O}\left(\frac{1}{\Lambda}\right) 
\end{equation} 
where $\Omega_{\Lambda} = [-\Lambda^{1/2}; \Lambda^{1/2}]^{d}$, 
and $1_{\Omega_{\Lambda}}$ is the characteristic function of
$\Omega_{\Lambda}$. 

Furthermore we have the following properties 
\begin{itemize} 
\item $T$ has a smooth, translation invariant kernel  
\item $T_{\Lambda, \varepsilon}$ and $T$ have high power decay  
\begin{equation}
  \exists n_{0} \mbox{ large, } \exists C_{T}(n_{0}) \mbox{ such that } 
    \forall (\Delta, \Delta'), \quad 
    \| 1_{\Delta} T_{\Lambda, \varepsilon} 1_{\Delta'}\| 
    \leqslant \frac{C_{T}(n_{0})}{\left[1 + L^{-1} d_{\Lambda}(\Delta,
    \Delta')\right]^{n_{0}}}
\end{equation}
and a similar relation for $T$ with $d_{\Lambda}$ being
replaced by $d$.  
\item $T(E, \lambda)$ is an analytic operator valued function of $E$ for
all $E$ in $]E_{1}, E_{2}[$ with a small $\lambda$-dependent radius of
analyticity. 
\item $T(E, \lambda)$ is a ${\cal C}^{\infty}$ operator-valued function of
$\lambda$ and admits an asymptotic expansion to all orders in $\lambda$. 
which is the formal perturbative expansion of 
\begin{equation}
  \int \! d\mu_{\xi}(V) \, e^{\frac{\mu_{0}^{2}}{2} <1, \xi^{-1} 1>} 
    \, e^{i \mu_{0} <V, \xi^{-1} 1>} \, \frac{1}{H_{0} -E  +
    \lambda V  -i (\mu+0^{+})} 
\end{equation}
($<>$ denotes the scalar product, {\it i.e.} 
$<f, Af> = \int \! \bar{f}(x) A(x, y) f(y)\, dx \, dy$)
\end{itemize}
\end{theorem}

This theorem is formulated in a rather general way so as to apply with 
minimum transformation to various situations (lattice or continuous models) 
and in any
dimension. Then we construct a concrete example with a two-dimensional
model. One can also refer to \cite{MPR} for a $d=3$ case. 

\subsection{Anderson model with an infra-red cut-off in dimension d=2}  
We consider 
\begin{equation}
  H = -\Delta_{\eta} + \lambda \eta_{E} V \eta_{E} 
\end{equation} 
where
\begin{itemize}
  \item $\Delta_{\eta}^{-1}$ is a ultra-violet regularized inverse Laplacian,
  {\it i.e.} there is a ${\cal C}_{0}^{\infty}$ function 
  $\eta_{\scriptscriptstyle UV}$ equal to 1 on ``low'' momenta such that  
\begin{equation}
  \Delta_{\eta}^{-1}(p) = \frac{\eta_{\scriptscriptstyle UV}(p)}{p^{2}}
\end{equation}
We will note $p^{2}$ instead of $- \Delta_{\eta}$, the UV-cutoff being then
 implicit. 
  \item we are interested in the mean Green's function for an energy $E=O(1)$ 
  \item $\eta_{E}$ is an infra-red cut-off which enforces 
\begin{equation}
  |p^{2} - E| \geqslant A \lambda^{2} |\log \lambda|^{2} 
\end{equation}
for some large constant $A$
  \item $V$ has covariance $\xi$ which is a ${\cal C}^{\infty}_{0}$
  approximation of a $\delta$-function 
\end{itemize}
This corresponds to the model away from the singularity $p^{2} = E$ in a 
multi-scale renormalization group analysis, we will show that it generates a 
small fraction of the expected imaginary part which is $O(\lambda^{2})$. 

Let $M^{1/2}$ be an even integer greater than 2, we define $j_{0}
\in \NN$ such that   
\begin{equation}
  M^{-j_{0}} \leqslant \inf_{\mbox{\scriptsize Supp}(\eta_{E})} |p^{2} - E| 
    \leqslant M^{-(j_{0}-1)}
\end{equation} 

Next, we construct a smooth partition of unity into cubes of side $M^{j_{0}}$ 
(they form a lattice $I\!\! D_{j_{0}}$) and we construct the fields 
$V_{\Delta}$'s accordingly. 

\begin{theorem}
\label{thprob2d}
\ \\ 
There exist constants $C_{0}$ and $C_{1}$ such that for any $\Lambda$, 
$a \geqslant 1$ and $\Delta \in I\!\! D_{j_{0}}$ we have 
\begin{equation}
  I\!\!P_{\Lambda} \left( \|D_{\Lambda, \mu} \eta_{E} V_{\Delta} \eta_{E} 
    D_{\Lambda, \mu}\| \geqslant a C_{0} j_{0} M^{j_{0}/2}\right) 
    \leqslant C_{1} e^{-\frac{1}{2} a^{2} M^{j_{0}/6}}  
\end{equation}
Furthermore theorem \ref{thRE} applies and $G_{E}$ is asymptotic to its
perturbative expansion so that it behaves more or less like 
\begin{equation}
  G_{E} \sim \frac{1}{p^{2} - E - i \eta_{E} O(\lambda^{2} |\log
  \lambda|^{-2}) \eta_{E}}
\end{equation} 
\end{theorem}

It is easy to extend this result to the case of a rotation invariant dispersion
relation and for energies outside the free spectrum not too close to the band
edge. In this case, the cut-off is no longer needed so that the result apply
to the full model.


\section{Resolvent polymer expansion with large field versus 
small field conditions}
\subsection{Sketch of proof for theorem {\protect \ref{thRE}}}
We give here the global strategy for proving theorem \ref{thRE}, the main
ingredient being the polymer expansion that we will detail in the following.
 
First we  recall (without proving them) 
some quite standard properties of gaussian measures. 
\begin{lemma}
\label{lemtrans}
Complex translation \\
Let $X$ be a gaussian random field with covariance $C$ and let $d\mu_{C}$ be
the associated measure. For any regular functional ${\cal F}(X)$ and any
function $f \in \mbox{Ran } C$, we have the following identity
\begin{equation}
  \int \! d\mu_{C}(X) \, {\cal F}(X) = e^{\frac{1}{2} <f, C^{-1} f>} \int \! 
    d\mu_{C}(X) \, {\cal F}(X-if) \, e^{i<X, C^{-1} f>}
\end{equation} 
\end{lemma} 

\begin{lemma}
\label{lemipp}
Integration by part \\
With the same notations than above we have
\begin{equation}
  \int \! d\mu_{C}(X) \, X(x) {\cal F}(X) = \int \! dy \, C(x, y) \int \! 
    d\mu_{C}(X) \frac{\delta}{\delta X(y)} {\cal F}(X)
\end{equation}
\end{lemma} 

Those  lemmas could for instance be easily proved for polynomial functionals
and extended through a density argument to a wide class of functionals. 
\hfill $\blacksquare$

Our starting point is obtained by applying lemma \ref{lemtrans} with
$\displaystyle f = \mu_{0} 1$. 
\begin{eqnarray}
\label{complextrans}
  G_{\Lambda, \varepsilon}(E+z, \lambda) &=& 
    \int \! d\mu_{\xi_{\Lambda}}(V) \, 
    e^{\frac{\mu_{0}^{2}}{2} <1, \xi_{\Lambda}^{-1} 1> +i \mu_{0} 
    <V, \xi_{\Lambda}^{-1} 1>} \, 
    \frac{1}{H_{0}^{(\Lambda)} - (E+ i\mu)  + 
    \lambda V - i\varepsilon - z} \\
  T_{\Lambda, \varepsilon}(E+z, \lambda) &=& 
    \int \! d\mu_{\xi_{\Lambda}}(V) \, 
    e^{\frac{\mu_{0}^{2}}{2} <1, \xi_{\Lambda}^{-1} 1> +i \mu_{0} 
    <V, \xi_{\Lambda}^{-1} 1>} \, 
    \frac{1}{U_{\Lambda, \mu}^{-1} +  
    \lambda D_{\Lambda, \mu} V D_{\Lambda, \mu} - (z + i\varepsilon)
    D_{\Lambda, \mu}^{2}} 
\end{eqnarray}

On one hand we earned something because now the resolvent operator in the
integral is bounded in norm independently of $\varepsilon$ (in the following
we will note $z$ instead of $z+ i\varepsilon$ and show convergence for any $z$
such that $|z| \ll \mu$, this would allow to prove analyticity in $z$).  
But on the other hand we have a huge
normalization factor to pay. However, we can remark that this normalization
factor is in fact equivalent to a factor $e$ per $L$-cube. 

Most of the demonstration amounts to a polymer expansion of $T_{\Lambda,
\varepsilon}$, {\it i.e.} we write $T_{\Lambda, \varepsilon}$ as a sum over
polymers of polymer activities 
\begin{eqnarray}
  T_{out, in} &=& \chi_{\Delta_{out}} T_{\Lambda, \varepsilon}
    \chi_{\Delta_{in}} \\ 
   T_{out, in} &=& \chi_{\Delta_{out}} \left[ U_{\Lambda, \mu} + 
     \frac{\lambda^{c_{1}}}{[1 + L^{-1} 
     d_{\Lambda}(\Delta_{in}, \Delta_{out})]^{n_{0}}} 
     \sum_{Y \in {\cal A}}  \lambda^{c_{2}|Y|} \Gamma_{Y} \, T(Y) \right] 
     \chi_{\Delta_{in}}
\end{eqnarray}
where $c_{1}$ and $c_{2}$ are small constants, $\Gamma_{Y}$ has decay in the
spatial extension of $Y$ and $\displaystyle \|\sum_{Y \in {\cal A}} T(Y)\|$ 
is bounded. Furthermore, $G(Y)$ is given by a functional integration over
fields $V_{\Delta}$'s corresponding to cubes in the support of the polymer
$Y$. This show that $T_{\Lambda, \varepsilon}$ is bounded and has a high power
decay uniformly in $\Lambda$. 

Next, when we consider $1_{\Omega_{\Lambda}} T_{\Lambda, \varepsilon}
1_{\Omega_{\Lambda}}$ we can divide the sum over polymers into a sum over
polymers with a large spatial extension (say $\Lambda^{2/3}$) and sum over
``small'' polymers. The large polymers will have a total contribution small
as $\Lambda^{-1}$ to some large power. For the small polymers, since we are
far away from the boundaries, their contribution calculated with
$d\mu_{\xi_{\Lambda}}$ will be equal to their contribution calculated with 
$d\mu_{\xi}$ up to a factor $\Lambda^{-n}$. In this way we can prove the
development (\ref{Lambdadev}). Smoothness of the kernel 
will be obtained because we will show that we can write 
\begin{equation}
  T(Y) = U_{\Lambda, \mu} \tilde{T}(Y) U_{\Lambda, \mu}
\end{equation} 
The convergence for any $z \ll \mu$ allows to show analyticity (we write
$z$-derivatives as Cauchy integrals so that we can show that they all exists
and do not grow too fast). 
Then an asymptotic expansion can be generated through the repeated use of
resolvent identity. 

Finally, for the density of states, we just need to remark that 
\begin{equation}
  G(0, 0) = \int \! dp \, dq \, G(p, q) = <\tilde{\delta}, G \, 
    \tilde{\delta}> 
\end{equation}
where $\tilde{\delta}$ is a regularized $\delta$-function because of the 
presence of the ultra-violet cut-off. Thus an asymptotic expansion for $G$ 
with respect to the operator norm will yield an asymptotic expansion for the
density of states.


\subsection{Improved polymer expansions}
Cluster expansions in constructive field theory lay heavily on a clever
application of the Taylor formula with integral remainder. Writing the full
Taylor series would amount to completely expand the perturbation series, which
most often diverges, and therefore should be avoided. A rather instructive
example of minimal convergent expansion is the Brydges-Kennedy forest formula:
you have a function defined on a set of links between pair of cubes
and you expand it not on all possible graphs but only on forests 
(cf \cite{Bry}).

For more complex objects a way to generalize such a formula can be
found in \cite{AR}, and we refer the reader to it for a more careful
treatment and for various proofs.  
Let us assume that we have a set of objects that we call monomers.
A sequence of monomers will be called a polymer, then we will expand 
a function defined on a set of monomers into a sum over 
allowed polymers. 

To be more precise, let ${\cal X}$ be a set of monomers, we define the set
${\cal Y}$ of polymers on ${\cal X}$ as the set of all finite sequences
(possibly empty) of elements of ${\cal X}$. Then a monomer can be identified
to a polymer of length 1. The empty sequence or empty polymer will be noted
$\emptyset$. We define on ${\cal Y}$
\begin{itemize}
  \item a concatenation operator: for $Y=(X_{1}, \ldots, X_{n})$ and 
$Y'=(X'_{1}, \ldots, X'_{n'})$, we define 
\begin{equation}
  Y \cup Y' = (X_{1}, \ldots, X_{n}, X'_{1}, \ldots, X'_{n'}) 
\end{equation}
  \item the notion of starting sequence: we say that $Y_{1}$ is a starting
sequence of $Y$ (equivalently that $Y$ is a continuation of $Y_{1}$) and we
note $Y_{1} \subset Y$ iff there exists $Y_{2}$ such that $Y = Y_{1} \cup
Y_{2}$ 
\end{itemize} 

Then we call allowed set (of polymers) any finite subset ${\cal A} \subset
{\cal Y}$ such that 
\begin{itemize}
  \item $\forall Y, Y' \quad Y' \subset Y \mbox{ and } Y \in {\cal A}
    \Rightarrow Y' \in {\cal A}$ 
  \item $\forall X, Y, Y' \quad Y \subset Y' \mbox{ and } Y\cup X \not \in
  {\cal A} \Rightarrow Y'\cup X \not \in {\cal A}$
\end{itemize}
the first condition implies that $\emptyset \in {\cal A}$ whenever ${\cal A}$
is non-empty. Finally, for $Y$ belonging to some allowed set ${\cal A}$, a
monomer $X$ is said to be admissible for $Y$ (according to ${\cal A}$) iff $Y
\cup X \in {\cal A}$.

\begin{lemma}
\label{lemclust}
\ \\
Let ${\cal X} = \{ X \}$ be a set of $N$ monomers and ${\cal Y}$ the set of
polymers on ${\cal X}$. We assume that we have an indexation of $\mbox{I}\!
\mbox{R}^{N}$ by ${\cal X}$, {\it i.e.} a bijection from ${\cal X}$ to $\{ 1,
\ldots, N \}$ so that an element of $\mbox{I}\! \mbox{R}^{N}$ can be noted 
$\vec{z} = (z_{X})_{X \in {\cal X}}$. 

For ${\cal F}$ a regular function from $\mbox{I}\! \mbox{R}^{N}$ to some
Banach space ${\cal B}$ and an allowed set ${\cal A} \subset {\cal Y}$,
the polymer expansion of ${\cal F}$ according to ${\cal A}$ is given through
the following identity 
\begin{eqnarray}
\label{eqclust0}
  {\cal F}(\vec{1}) &\!\!\! \equiv& \!\!\! {\cal F}(1, \ldots, 1) \nonumber \\ 
    &\!\!\!=& \!\!\! \sum_{n \geqslant 0} \, 
    \sum_{Y=(X_{1}, \ldots, X_{n}) \in
    {\cal A}}  \int_{\lefteqn{\scriptstyle 1>h_{1}> \ldots h_{n} >0}} 
    \quad dh_{1} \ldots dh_{n}  
    \left( \prod_{X \in Y} \frac{\partial}{\partial z_{X}} \right)
    {\cal F}\left[\vec{z}(Y, \{h_{i}\})\right]
\end{eqnarray}
where $\vec{z}(Y,\{h_{i}\})$ is given by
\begin{equation}
  z_{X}(Y,\{h_{i}\}) = \left\{
  \begin{array}{ll}
    0&\mbox{if $X$ is admissible for $Y$} \\ 
    1&\mbox{if $X$ is not admissible for $\emptyset$} \\
    h_{i} &\mbox{if $X$ not admissible for $Y$ and $X=X_{j}$ for some $j$,} 
     \\
     & \mbox{in which case } i = \max \{j / X = X_{j} \} \\
    h_{i}&\mbox{with } i = \min \{j / X \mbox{ not admissible for } 
    (X_{1}, \ldots X_{j}) \}, \mbox{ otherwise}
  \end{array}
  \right.
\end{equation}
\end{lemma}

\noindent {\it Proof} \\
The proof is made through an inductive iteration of a first order Taylor
formula. We start with ${\cal F}(\vec{1})$ and put a common interpolating
parameter $h_{1}$ on all admissible monomers for the empty set, i.e. we make a
first order Taylor expansion with integral remainder of ${\cal F} \left[ h_{1}
\vec{z}_{1} + (\vec{1} - \vec{z}_{1}) \right]$ between 0 and 1, with
$\vec{z}_{1}$ being the vector with entries 1 or 0 according to whether the
corresponding monomer is admissible or not. Then each partial derivative
acting on ${\cal F}$ can be seen as taking down the corresponding monomer so
that terms can be seen as growing polymers. 
The iteration goes as follow: 
for a term of order $n$ corresponding to a given polymer $Y$ and having $n$
interpolating parameters $1> h_{0} > \ldots > h_{n} >0$ we put a common 
parameter $h_{n+1}$ interpolating between $0$ and $h_{n}$ on all monomers
admissible for $Y$. It is easy to check that the process is finite since ${\cal
A}$ is finite and that one obtains the desired formula. 
\hfill $\blacksquare$

In the following our monomers are sets of cubes (that we call the support of
the monomer) and links between those cubes. When we take down a polymer, 
we connect all the cubes in its support and maybe some more cubes. 
Thus a polymer is made of several connected regions, we will say that it 
is connected if it has a single connected component. 
The rules of admissibility will be to never
take down a monomer whose support is totally contained in a connected
region. 

In this case, one can show that the interpolating parameters depend only of
the connected component to which the corresponding monomer belongs so that one
can think to ``factorize'' the connected components. We define generalized
polymers as sets of connected polymers. Then a generalized polymer $Y = \{
Y_{1}, \ldots , Y_{p} \}$ is allowed if the polymer $Y_{1} \cup \ldots \cup
Y_{p}$ is allowed (this does not depend of the order of the $Y_{i}'s)$. 
Equation (\ref{eqclust0}) becomes
\begin{equation}
\label{eqclust}
  {\cal F}(\vec{1}) = \sum_{{Y=\{ Y_{1}, \ldots, Y_{p} \}} \atop 
    {Y_{i} = (X_{i}^{1}, \ldots X_{i}^{n_{i}})}} \, 
    \left( \prod_{i=1}^{p} \int_{\lefteqn{\scriptstyle 1>h_{i}^{1}> \ldots
    h_{i}^{n_{i}} >0}} \quad dh^{1}_{i} \ldots dh^{n_{i}}_{i} \right) \, 
    \left( \prod_{X \in Y} \frac{\partial}{\partial z_{X}} \right)
    {\cal F}\left[\vec{z}(Y, \{ h_{i}^{j} \})\right]
\end{equation}
where the sum extends on all allowed generalized polymers, and 
$\vec{z}(Y,\{ h_{i}^{j} \})$ is given by
\begin{equation}
\label{paramclust}
  z_{X}(Y,\{ h_{i}^{j} \}) = \left\{
  \begin{array}{ll}
    0&\mbox{if $X$ is admissible for $Y$, {\it i.e.} for } Y_{1} \cup \ldots 
    \cup Y_{p} \\ 
    1&\mbox{if $X$ is not admissible for $\emptyset$} \\
    h_{i}^{j} &\mbox{if $X=X_{i}^{j}$ for some $i$ and $j$} \\
    h_{i}^{j} &\mbox{where $X$ is not admissible for $Y_{i}$ and } \nonumber \\
    & j= \min \{k / X \mbox{ not admissible for } (X_{i}^{1}, 
    \ldots, X_{i}^{k})\}, \mbox{ otherwise}
  \end{array}
  \right.
\end{equation}


\subsection{Large/small field decomposition}
Semi-perturbative expansion (like cluster expansions) are convergent 
only when the ``perturbation'' is small (in our
case the operators $V_{\Delta}$'s). Thus it
is very important to distinguish between the so called {\it small field
regions} where perturbations will work and the {\it large field regions} where
we must find other estimates (they will come mostly from the exponentially
small probabilistic factor attached to those regions).

We take a ${\cal C}^{\infty}_{0}$ function
$\varepsilon$ such that 
\begin{itemize}
\item $0 \leqslant \varepsilon \leqslant 1$ 
\item $\mbox{Supp}(\varepsilon) \subset [0, 2]$ 
\item $\varepsilon_{|_{[0, 1]}} = 1$
\end{itemize}
Then for each $\Delta$ 
we define 
\begin{equation}
  \varepsilon_{\Delta} (V_{\Delta}) = \varepsilon \left( 
  \frac{\|D_{\Lambda, \mu} V_{\vec{\Delta}} D_{\Lambda, \mu}\|}{a 
    \lambda^{-1/4} C_{0}} \right) \quad 
    \mbox{and} \quad \eta_{\Delta}= 1-\varepsilon_{\Delta}
\end{equation}
where $a = O(1)$. Then we can expand 
\begin{equation}
\label{lsdecomp}
  1 = \prod_{\Delta} (\varepsilon_{\Delta} + \eta_{\Delta}) = \sum_{N
    \geqslant 0} \, \sum_{\Omega = \{ \Delta_{1}, \ldots, \Delta_{N}\}} 
    \left(\prod_{\Delta \in \Omega} \eta_{\Delta}\right) 
    \left(\prod_{\Delta \not \in \Omega} \varepsilon_{\Delta}\right) 
\end{equation}
where $\Omega$ is the large field region whose contribution will be 
isolated through the following lemma.   

\begin{lemma}
\label{lemlargef}
\ \\
Let $\Omega$ be a large field region made of $N$ cubes
$\Delta_{1}$, \ldots, $\Delta_{N}$ and  $A$ any operator such
that 
\begin{equation}
  \forall {D} \subset \{1, \ldots N \}, \quad \mbox{$\displaystyle A + 
    \sum_{i \in {D}} B_{i}$ is invertible} 
\end{equation}
($B_{i}$ stands for $B_{\Delta_{i}} \equiv \lambda D_{\Lambda, \mu}
V_{\Delta_{i}} D_{\Lambda, \mu}$).  

We have the following identity 
\begin{equation}
\label{eqlargef}
  \frac{1}{\displaystyle A + \sum B_{i}} = \sum_{n=0}^{N} \, (-1)^{n} 
    \hspace{-1em} 
    \sum_{i_{1} \in \{1 \ldots N\}}
    \sum_{{i_{2} \in \{1 \ldots N\}} \atop {i_{2} \not \in \{i_{1}\}}} \ldots
    \sum_{{i_{n} \in \{1 \ldots N\}} \atop {i_{n} \not \in \{i_{1} \ldots
    i_{n-1}\}}} \frac{1}{A} O_{n} \frac{1}{A} \ldots O_{1} \frac{1}{A}
\end{equation} 
where
\begin{equation}
  O_{p} = B_{p} - \left(\sum_{i \in \{1 \ldots p\}} B_{i} \right)
    \frac{1}{\displaystyle A + \sum_{i \in \{1 \ldots p\}} B_{i}} B_{p} 
\end{equation} 
\end{lemma}

\noindent {\it Proof} \\
The proof relies on resolvent expansion identities
\begin{equation}
  \frac{1}{A+B} = \frac{1}{A} \left(I - B \frac{1}{A+B}\right) = 
    \left(I - \frac{1}{A+B} B\right) \frac{1}{A}
\end{equation}
We show by induction that for all $m \in \{1, \ldots, N\}$ we have 
\begin{eqnarray}
  \frac{1}{A + \sum B_{i}}  &=& \sum_{n=0}^{m-1} \, (-1)^{n} \hspace{-1em} 
    \sum_{{(i_{1}, \ldots i_{n})} \atop {i_{k} \not \in
\{i_{1} \ldots i_{k-1}\}}} \frac{1}{A} O_{n} \frac{1}{A} \ldots O_{1}
\frac{1}{A} + (-1)^{m} R_{m} \\  
  R_{m} &=& \hspace{-1em} \sum_{{(i_{1}, \ldots i_{m})} \atop {i_{k} \not \in
\{i_{1} \ldots i_{k-1}\}}} \frac{1}{A + \sum B_{i}} B_{i_{m}} \frac{1}{A} 
O_{m-1} \frac{1}{A} \ldots O_{1} \frac{1}{A} 
\end{eqnarray}

The case $m=1$ is obtained by a resolvent expansion 
\begin{equation}
\frac{1}{A + \sum B_{i}} = \frac{1}{A} - \sum_{i_{1}} \frac{1}{A + \sum B_{i}}
 B_{i_{1}} \frac{1}{A} 
\end{equation}

Then we go from $m$ to $m+1$ with 2 steps of resolvent expansion. We write 
\begin{eqnarray}
\frac{1}{A + \sum B_{i}} &=& \left(I - \hspace{-1em} \sum_{i_{m+1} \not \in
\{i_{1} \ldots i_{m}\}} \frac{1}{A + \sum B_{i}} B_{i_{m+1}}\right)  
\frac{1}{\displaystyle A + \sum_{k=1}^{m} B_{i_{k}}}  \\ 
&=& \left(I - \hspace{-1em} \sum_{i_{m+1} \not \in \{i_{1} \ldots i_{m}\}}
\frac{1}{A + \sum B_{i}} B_{i_{m+1}} \right)  \, \frac{1}{A} 
\left(I - \sum_{k=1}^{m} B_{i_{k}}  
\frac{1}{\displaystyle A + \sum_{l=1}^{m} B_{i_{l}}} \right)
\end{eqnarray} 

Finally, for $m=N$ we make a last resolvent expansion on the rest term $R_{N}$
by writing 
\begin{equation}
\frac{1}{A + \sum B_{i}} =   \frac{1}{A} \left(I - \sum B_{i} 
\frac{1}{A+\sum B_{i}} \right)
\end{equation}
\hfill $\blacksquare$ 

If we look at 
\begin{equation}
  \chi_{\Delta_{out}} \frac{1}{A+\sum B_{i}} \chi_{\Delta_{in}}
\end{equation}
and fix $\{\Delta_{i_{1}}, \ldots \Delta_{i_{n}}\}$, we can see that summing
over the sequences $(i_{1}, \ldots i_{n})$ and choosing a particular term for
each $O_{p}$ amounts to construct a tree on 
$\{ \Delta_{in}, \Delta_{out}, \Delta_{i_{1}}, \ldots \Delta_{i_{n}}\}$. 

We define an oriented link $l$ as a couple of cubes that we note $(l.y, l.x)$,
then $\vec{\cal L}$ is the set of oriented links. Given two cubes
$\Delta_{in}$ and $\Delta_{out}$ and a set of cubes $\Omega = \{ \Delta_{1},
\ldots \Delta_{n}\}$ we construct the set ${\cal T}_{R}(\Delta_{in},
\Delta_{out}, \Omega)$ of oriented trees going form
$\Delta_{in}$ to $\Delta_{out}$ through $\Omega$ as the sequences 
$(l_{1}, \ldots l_{n+1}) \in \vec{\cal L}^{n+1}$ which satisfy 
\begin{itemize}
  \item $l_{1}.x = \Delta_{in}$ 
  \item $l_{n+1}.y = \Delta_{out}$
  \item $\forall k \in \{1, \ldots n\}, \, l_{k}.y \in \Omega $
  \item $\forall k \in \{2, \ldots n+1\}, \, l_{k}.x \in \{ l_{1}.y, \ldots
    l_{k-1}.y \} $
  \item $\forall k \in \{2, \ldots n\}, \, l_{k}.y \not \in \{ l_{1}.y, \ldots
    l_{k-1}.y \} $  
\end{itemize}
Then we have the following equivalent formulation of lemma \ref{lemlargef}.  
\begin{lemma}
\label{ldecoupling}
\ \\
Let $\Omega$ be a large field region made of $N$ cubes
$\Delta_{1}$, \ldots, $\Delta_{N}$ and  $A$ any operator such
that 
\begin{equation}
  \forall {D} \subset \{1, \ldots N \}, \quad \mbox{$\displaystyle A + 
    \sum_{i \in {D}} B_{i}$ is invertible} 
\end{equation}

We have the following identity 
\begin{eqnarray}
\label{eqldecoupling}
  \chi_{\Delta'} \frac{1}{\displaystyle A + \sum B_{i}}
\chi_{\Delta} &=& \sum_{n=0}^{N} \, (-1)^{n} \hspace{-1.5em}
\sum_{{\Omega' \subset \Omega} \atop {\Omega' = \{\Delta'_{1}, \ldots
\Delta'_{n}\}}} 
\sum_{{{\cal T} \in {\cal T}_{R}(\Delta, \Delta', \Omega')} 
\atop {{\cal T} = (l_{1}, \ldots l_{n+1})}}  \nonumber \\ 
&& \quad \chi_{\Delta'} \frac{1}{A}
O_{n}(l_{n+1}.x, l_{n}.y) \frac{1}{A} \ldots O_{1}(l_{2.x}, l_{1}.y) 
\frac{1}{A} \chi_{\Delta}
\end{eqnarray} 
where
\begin{equation}
  O_{p}(\Delta_{j}, \Delta_{i}) = B_{\Delta_{i}} \delta_{\Delta_{i}
\Delta_{j}} -  
B_{\Delta_{j}} \frac{1}{\displaystyle A + \sum_{i \in \{1 \ldots p\}} B_{i}} 
B_{\Delta_{i}} 
\end{equation} 
\end{lemma}
The proof being just a rewriting of lemma \ref{lemlargef} is quite immediate. 
\hfill $\blacksquare$

Thanks to this lemma we can factorize out the contribution of the large field
region, then we need to extract spatial decay for the resolvent in the 
small field region. However a kind of Combes-Thomas estimate (\cite{CT}) 
would not be enough because of the normalization factor that we must pay. 
For this reason, we will make a polymer expansion to determine which region 
really contributes to the resolvent.


\subsection{Polymer expansion for the resolvent in the small field region} 

For some large field region $\Omega$, we want to prove the decay of 
\begin{equation}
  \frac{1}{\displaystyle U_{\Lambda, \mu}^{-1} + \lambda \sum_{\Delta
    \not \in \Omega} D_{\Lambda, \mu} V_{\Delta} D_{\Lambda, \mu} - z
    D_{\Lambda, \mu}^{2}} \equiv R U_{\Lambda, \mu}  
\end{equation}
and get something to pay for the normalization factor. 

We define the set ${\cal L}$ of links as the set of pair of cubes, and 
${\cal L}(\Omega)$ as the set of links which do not connect two cubes of
$\Omega$. Then for $l = \{\Delta, \Delta'\}$ we define
\begin{eqnarray}
  Q_{l} &=& \lambda (\chi_{\Delta} U_{\Lambda, \mu} D_{\Lambda, \mu} 
    V_{\Delta'} D_{\Lambda, \mu}+
    \chi_{\Delta'} U_{\Lambda, \mu} D_{\Lambda, \mu} V_{\Delta} 
    D_{\Lambda, \mu}) 
    - z (\chi_{\Delta} C_{\Lambda, \mu} \chi_{\Delta'} 
    + \chi_{\Delta'} C_{\Lambda, \mu} \chi_{\Delta} ) \\
  R_{l} &=& \chi_{\Delta} R \chi_{\Delta'} + \chi_{\Delta'} R 
    \chi_{\Delta} \\ 
  U_{l} &=& \chi_{\Delta} U_{\Lambda, \mu} \chi_{\Delta'} + 
    \chi_{\Delta'} U_{\Lambda, \mu} \chi_{\Delta} 
\end{eqnarray} 
with the convention that $V_{\Delta} = 0$ if
$\Delta \in  \Omega$.  

For any fixed $l_{0} = (\Delta_{0}, \Delta_{0}')$ we expand $R_{l_{0}}$ 
on ${\cal L}(\Omega)$ with the rule that for any growing polymer 
\begin{itemize}
\item if we have two adjacent connected components $Y_{1}$ and $Y_{2}$ 
(such that $d_{\Lambda}(Y_{1}, Y_{2}) = 0$) we connect the two components 
\item we connect $\Delta_{0}$ (resp. $\Delta_{0}'$) to any adjacent polymer 
component 
\end{itemize} 
This allows to take into account that the operators localized on a pair of 
cubes have their support extending to the neighboring cubes. 

Let us notice that if $A$ and $B$ have disjoint support, we have 
\begin{equation}
  \frac{1}{I+A+B} = \frac{1}{I+A} \, \frac{1}{I+B}
\end{equation}

Then it is easy to see that the expansion of $R_{l_{0}}$ involves only totally
connected polymers which connect $\Delta_{0}$ to $\Delta_{0}'$, 
because the other terms necessarily contain a product of two operators
with disjoint supports which gives zero. We note ${\cal A}(\Omega, l_{0})$ the
corresponding set of polymers which is a decreasing function of $\Omega$, {\it
i.e.}
\begin{equation} 
\Omega' \subset \Omega \Rightarrow {\cal A}(\Omega, l_{0}) \subset
{\cal A}(\Omega', l_{0})
\end{equation}  
Then, according to (\ref{eqclust}), our expansion looks like 

\begin{eqnarray}
  R_{l_{0}}(\Omega) &=& \sum_{n \geqslant 0} \,  \sum_{{Y \in {\cal A}(\Omega,
    l_{0})} \atop {Y = (X_{1}, \ldots, X_{n})}} 
    \int_{1>h_{1}> \ldots h_{n}>0} \hspace{-4em} dh_{1} \, \ldots \,
    dh_{n} \left(\prod_{X \in Y} \frac{\partial }{\partial z_{X}}\right)  
    \frac{1}{\displaystyle I + \sum_{X \in {\cal L}(\Omega)} z_{X} Q_{X}} 
    \left[\vec{z}(Y, \{h_{i}\})\right] \\ 
  &=& \sum_{Y \in {\cal A}(\Omega, l_{0})} \int \prod_{i} dh_{i} 
   \left(\prod_{X \in Y} \frac{\partial }{\partial u_{X}}\right)  
   \frac{1}{\displaystyle I + \sum_{X \in {\cal L}(\Omega)} z_{X} Q_{X} 
   + \sum_{X \in Y} u_{X} Q_{X}}  
    \left[\vec{z}(Y, \{h_{i}\}), \vec{0}\right] 
\end{eqnarray}
Then in the second expression, we rewrite the derivatives as Cauchy integrals
so that 
\begin{eqnarray}
\label{eqresexp}
  R_{l_{0}}(\Omega) &=& \sum_{Y \in {\cal A}(\Omega, l_{0})} \int \prod_{i} 
    dh_{i} \, \left(\prod_{X \in Y} \oint \frac{du_{X} }{2i\pi 
    u_{X}^{2}}\right) 
    \frac{1}{\displaystyle I + \sum_{X \in {\cal L}(\Omega)} 
    z_{X}(Y, \{h_{i}\}) Q_{X}
    +\sum_{X \in Y} u_{X} Q_{X}} \\ 
\label{Yexpan}
  && \equiv \sum_{Y \in {\cal A}(\Omega, l_{0})} R(Y)  = I +  
\sum_{Y \in {\cal A}^{*}(\Omega, l_{0})} R(Y)
\end{eqnarray}
where ${\cal A}^{*}(\Omega, l_{0}) = {\cal A}(\Omega, l_{0}) / \{
\emptyset\}$

We suppose that we fixed $n_{1}$ the power rate of decay in
$d_{\Lambda}(\Delta_{1}, \Delta_{3})$ of
$\|\chi_{\Delta_{1}} D_{\Lambda, \mu} V_{\Delta_{2}} D_{\Lambda, \mu}
\chi_{\Delta_{3}}\|$ and $\| \chi_{\Delta_{1}} C_{\Lambda, \mu}
\chi_{\Delta_{3}}\|$, then we have the following lemma 
\begin{lemma}
\label{lemresexp}
\ \\
For $n_{2} = n_{1} - 3(d+1)$ and $\lambda$ small enough, we have 
\begin{eqnarray}
  && \forall l_{0}=\{\Delta_{0}, \Delta_{0}'\}, \, \forall Y \in 
  {\cal A}^{*}(\emptyset, l_{0}) \equiv {\cal A}^{*}(l_{0}), \nonumber \\ 
  && \quad \quad \quad
  \| R(Y) \| \leqslant \frac{\lambda^{|Y|/4}}{[1 + L^{-1}
  d_{\Lambda}(\Delta_{0}, \Delta_{0'})]^{n_{2}}} \Gamma(Y) 
  \mbox{ with} \sum_{Y \in {\cal A}^{*}(l_{0})} \Gamma(Y) \leqslant 1
\end{eqnarray}
where $|Y|$ is the number of monomers in $Y$.
\end{lemma} 

\noindent {\it Proof} \\ 
Since we are in the small field region 
\begin{equation}
  \forall l=\{\Delta, \Delta'\}, \quad \| Q_{l}\| \leqslant \frac{ O(1) \, 
 \lambda^{3/4}}{[1+L^{-1} d_{\Lambda}(\Delta, \Delta')]^{n_{1}-(d+1)}} 
\end{equation}
Then in (\ref{eqresexp}) we can integrate each $u_{l}$ on a circle of radius 
\begin{equation}
  R_{l} = \lambda^{-1/2} [1+L^{-1} d_{\Lambda}(\Delta, \Delta')]^{n_{1} -
    2 (d+1)} 
\end{equation}
while staying in the domain of analyticity for $u_{l}$ and have a resolvent
bounded in norm by say $2$ (if $\lambda$ is small enough). Thus 
\begin{eqnarray}
  \| R(Y) \| &\leqslant& \int_{1>h_{1}> \ldots h_{|Y|}>0} \hspace{-4em} dh_{1} 
\, \ldots \, dh_{|Y|} \, \left(\prod_{{X \in Y} \atop {X=\{\Delta_{X},
\Delta_{X}'\}}} \frac{O(1) \, \lambda^{1/2}}{[1+L^{-1} 
d_{\Lambda}(\Delta_{X}, \Delta_{X}')]^{n_{1} - 2(d+1)}} \right) \\ 
  &\leqslant& \frac{\lambda^{|Y|/4}}{[1 + L^{-1}
d_{\Lambda}(\Delta_{0}, \Delta_{0'})]^{n_{2}}} \, \frac{\left[O(1)
\lambda^{1/4}\right]^{|Y|}}{|Y|!} 
\left(\prod_{{X \in Y} \atop {X=\{\Delta_{X}, \Delta_{X}'\}}} 
\frac{1}{[1+L^{-1} d_{\Lambda}(\Delta_{X}, \Delta_{X}')]^{d+1}} \right) 
\end{eqnarray}
this demonstrates the first part of the lemma with 
\begin{equation}
  \Gamma(Y) = \frac{\left[O(1) \lambda^{1/4}\right]^{|Y|}}{|Y|!}  
\left(\prod_{X \in Y} \Gamma_{X} \right) \quad \mbox{ and } \sum_{X \ni
\Delta} \Gamma_{X} = O(1) 
\end{equation} 

A link $l=\{\Delta, \Delta'\}$ can either be a true link when $\Delta \neq
\Delta'$ or a tadpole when both cubes collapse. Our expansion rules insure
that there is at most 1 tadpole per cube of $Supp(Y)$ the support of $Y$. If
we forget about proximity links (that we connect also adjacent cubes) then a
polymer with $m$ true links and $p$ tadpoles has a support of $m+1$ cubes (2
of them being $\Delta_{0}$ and $\Delta_{0}$') and the true links make a tree
on the support of $Y$. If we take into account the proximity links then two
connected links in the tree on $Y$ are adjacent instead of sharing a
common cube, we will forget about this since it would induce at most a
factor $O(1)^{|Y|}$. The links in $Y$ are ordered, but we can take them
unordered by eating up the $1/|Y|!$ we have in $\Gamma(Y)$. 

Then the sum over $Y$ can be decomposed as 
\begin{itemize}
  \item choose $m \geqslant 1$
  \item choose $m-1$ cubes $\{ \Delta_{1}, \ldots, \Delta_{m-1}\}$ 
  \item chose a tree ${\cal T}$ 
  on $\{ \Delta_{0}, \Delta_{0}', \Delta_{1}, \ldots, \Delta_{m-1}\}$
  \item choose $0 \leqslant p \leqslant m+1$ 
  \item place $p$ tad-poles on $\{ \Delta_{0}, \Delta_{0}', \Delta_{1},
    \ldots, \Delta_{m-1}\}$
\end{itemize} 

We can perform the sum on tadpole configurations because for $p$ tadpoles, we
have a factor $\left[O(1) \lambda^{1/4}\right]^{p}$ coming from the tadpoles
and at most $\left({m+1} \atop {p}\right)$ configurations. Thus 
\begin{eqnarray}
  \sum_{Y \in {\cal A}^{*}(l_{0})} \!\!\!\! \Gamma(Y) &\leqslant& 
  \sum_{m\geqslant 1} \sum_{\{ \Delta_{1}, \ldots \Delta_{m-1}\}} 
\sum_{{\cal T}} \left[1 + O(1) \lambda^{1/4}\right]^{m+1} \nonumber \\ 
&& \quad 
\left[O(1) \lambda^{1/4}\right]^{m} \left( 
\prod_{X \in {\cal T}} \Gamma_{X}\right)
\end{eqnarray}
Then we fix first the form of ${\cal T}$ then we sum over the positions of
$\Delta_{1}$, \ldots, $\Delta_{m-1}$. But since the cubes are now labeled we
get $(m-1)!$ the desired sum.
\begin{equation}
  \sum_{Y \in {\cal A}^{*}(l_{0})} \Gamma(Y) \leqslant \sum_{m\geqslant 1} 
\left[O(1) \lambda^{1/4}\right]^{m} \frac{1}{(m-1)!} 
\sum_{\cal T} \sum_{(\Delta_{1}, \ldots, \Delta_{m-1})} 
\left( \prod_{X \in {\cal T}} \Gamma_{X}\right)
\end{equation}
We choose $\Delta_{0}$ as the root of our tree and suppose that the position
of $\Delta_{0}'$ is not fixed. Then the sum over the position
of the cubes is made starting from the leaves thanks to the decaying factors
$\Gamma_{X}$ (cf. \cite{Riv2}), this costs a factor $O(1)^{m}$. 

Finally, the sum over ${\cal T}$, which is a sum
over unordered trees, is performed using Cayley's theorem which states that
there are $(m+1)^{m-1}$ such trees.  
\begin{equation}
  \sum_{Y \in {\cal A}^{*}(l_{0})} \Gamma(Y) \leqslant \sum_{m\geqslant 1} 
\left[O(1) \lambda^{1/4}\right]^{m} \frac{(m+1)^{m-1}}{(m-1)!} \leqslant O(1)
\lambda^{1/4} \leqslant 1  
\end{equation}
for $\lambda$ small enough. 
\hfill $\blacksquare$

We note that we can perform the same expansion on 
\begin{equation}
  R'= U_{\Lambda, \mu}^{-1} \frac{1}{\displaystyle U_{\Lambda, \mu}^{-1} + 
    \lambda \sum_{\Delta
    \not \in \Omega} D_{\Lambda, \mu} V_{\Delta} D_{\Lambda, \mu} - z
    D_{\Lambda, \mu}^{2}}  
\end{equation}


\subsection{Summation and bonds on $T$}
We define 
\begin{equation}
  T_{out, in} = \chi_{\Delta_{out}} T_{\Lambda, \varepsilon}
\chi_{\Delta_{in}} 
\end{equation}
We can combine equations (\ref{complextrans}), (\ref{lsdecomp}),
(\ref{eqldecoupling}) and (\ref{Yexpan}) to write 
\begin{eqnarray}
  \lefteqn{T_{out, in}} && \quad \quad = \int \! \otimes 
d\mu_{\xi_{\Lambda}^{\Delta}} 
(V_{\Delta}) \, e^{ \frac{\mu_{0}^{2}}{2} < 1, \xi_{\Lambda}^{-1} 1> + i
\mu_{0} <V, \xi_{\Lambda}^{-1} 1>} \, 
\sum_{N \geqslant 0} \, \sum_{\Omega = \{ \Delta_{1}, \ldots \Delta_{N} \}} 
\left(\prod_{\Delta \in \Omega} \eta_{\Delta}\right) \, 
\left(\prod_{\Delta \not \in \Omega} \varepsilon_{\Delta} \right) 
\nonumber \\
&& \quad \quad \quad \sum_{n=0}^{N} (-1)^{n}  
\sum_{{\Omega' \subset \Omega} \atop {|\Omega'| = n}}  
\sum_{{{\cal T} \in {\cal T}_{R}(\Delta_{in}, \Delta_{out}, \Omega')} 
\atop {{\cal T} = (l_{1}, \ldots l_{n+1})}}  
\sum_{{(\Delta_{2}^{x}, \ldots \Delta_{n+1}^{x})} \atop {(\Delta_{1}^{y}, 
\ldots \Delta_{n}^{y})}} 
\sum_{(\Delta_{1}^{z}, \ldots \Delta_{n+1}^{z})}  
\sum_{{(Y_{1}, \ldots Y_{n+1})} \atop {Y_{i} \in {\cal A}(\Omega, 
\{\Delta_{i}^{y}, \Delta_{i}^{z}\})}} \nonumber \\
&& U_{\Delta_{out}, \Delta_{n+1}^{z}} R'_{\Delta_{n+1}^{z},
\Delta_{n+1}^{x}} T_{n}(l_{n+1}.x, l_{n}.y) 
R_{\Delta_{n}^{y}, \Delta_{n}^{z}} U_{\Delta_{n}^{z}, \Delta_{n}^{x}} 
\ldots T_{1}(l_{2.x}, l_{1}.y) R_{\Delta_{1}^{y}, \Delta_{1}^{z}}
  U_{\Delta_{1}^{z}, \Delta_{in}}
\end{eqnarray}
where we pretend that the $\chi_{\Delta}$'s are sharp otherwise we would have
to deal with adjacent cubes but it's an irrelevant complication. Furthermore,
for the leftmost term we made a polymer expansion of $U_{\Lambda, \mu} R'$ 
instead of $RU_{\Lambda, \mu}$ so that we can write $T_{out, in}$ as 
\begin{equation}
  T_{out, in} = \chi_{\Delta_{out}} \left(U_{\Lambda, \mu} + U_{\Lambda, \mu}
\tilde{T} U_{\Lambda, \mu} \right) \chi_{\Delta_{in}}
\end{equation}

The crucial point here is to notice that for any cube $\Delta$, each term
where $\Delta$ appears in $\Omega$ but not in $\Omega'$ pairs with a
corresponding term where $\Delta \not \in \Omega$ and $\Delta \not \in
\bigcup \mbox{Supp}(Y_{i})$ ({\it i.e.} $\Delta$ has been killed in every
polymer expansion). Then the corresponding $\varepsilon_{\Delta}$ and
$\eta_{\Delta}$ add up back to 1 so that 

\begin{eqnarray}
  \lefteqn{T_{out, in}} && \quad \quad = \sum_{n \geqslant 0} (-1)^{n}  
\sum_{\Omega = \{ \Delta_{1}, \ldots \Delta_{n} \}}   
\sum_{{{\cal T} \in {\cal T}_{R}(\Delta_{in}, \Delta_{out}, \Omega)} 
\atop {{\cal T} = (l_{1}, \ldots l_{n+1})}} 
\sum_{{(\Delta_{2}^{x}, \ldots \Delta_{n+1}^{x})} \atop {(\Delta_{1}^{y}, 
\ldots \Delta_{n}^{y})}} \sum_{(\Delta_{1}^{z}, \ldots \Delta_{n+1}^{z})} 
\sum_{{(Y_{1}, \ldots Y_{n+1})} \atop {Y_{i} \in {\cal A}(\Omega, 
\{\Delta_{i}^{y}, \Delta_{i}^{z}\})}} 
\nonumber \\ 
&& \quad \quad \int \! \otimes d\mu_{\xi_{\Lambda}^{\Delta}} (V_{\Delta}) 
\, e^{ \frac{\mu_{0}^{2}}{2} < 1, \xi_{\Lambda}^{-1} 1> + i \mu_{0} <V,
\xi_{\Lambda}^{-1} 1>} \, 
\left(\prod_{\Delta \in \Omega} \eta_{\Delta}\right) 
\left(\prod_{\Delta \in \cup \mbox{\footnotesize Supp}(Y_{i})} 
\varepsilon_{\Delta} \right) 
\nonumber \\
&& U_{\Delta_{out}, \Delta_{n+1}^{z}} R'_{\Delta_{n+1}^{z},
\Delta_{n+1}^{x}} T_{n}(l_{n+1}.x, l_{n}.y) 
R_{\Delta_{n}^{y}, \Delta_{n}^{z}}
U_{\Delta_{n}^{z}, \Delta_{n}^{x}} 
\ldots T_{1}(l_{2.x}, l_{1}.y) R_{\Delta_{1}^{y}, \Delta_{1}^{z}} 
U_{\Delta_{1}^{z}, \Delta_{in}}
\end{eqnarray}

The factor $e^{ \frac{\mu_{0}^{2}}{2} < 1, \xi_{\Lambda}^{-1} 1> + i \mu_{0}
<V, \xi_{\Lambda}^{-1} 1>}$ correponds to the translation of $V$ by 
$-i\mu_{0}$, this is equivalent to have translated all the $V_{\Delta}$'s 
by $-i \mu_{0} \chi_{\Delta}$ therefore we can write it as
\begin{equation}
  \prod_{\Delta} e^{ \frac{\mu_{0}^{2}}{2} < \chi_{\Delta},
(\xi_{\Lambda}^{\Delta})^{-1} \chi_{\Delta}> + i \mu_{0}
<V_{\Delta}, (\xi_{\Lambda}^{\Delta})^{-1} \chi_{\Delta}>}
\end{equation} 
then we can perform the integration on all $V_{\Delta} \not \in \Omega \cup
(\bigcup \mbox{Supp}(Y_{i}))$ so that the normalization factor reduces to 
\begin{equation}
  \prod_{\Delta \in \Omega \cup (\cup \mbox{\footnotesize Supp}(Y_{i}))} 
e^{ \frac{\mu_{0}^{2}}{2} < \chi_{\Delta},
(\xi_{\Lambda}^{\Delta})^{-1} \chi_{\Delta}> + i \mu_{0}
<V_{\Delta}, (\xi_{\Lambda}^{\Delta})^{-1} \chi_{\Delta}>}
\end{equation}
This amounts to pay a constant per cube of $\Omega \cup 
(\bigcup \mbox{Supp}(Y_{i}))$, this is done in $\Omega$ with a fraction of the
probabilistic factor coming from the large field condition and in $\bigcup
\mbox{Supp}(Y_{i})$ with a fraction of the factor $\lambda^{\sum |Y_{i}|/4}$
coming from the $R's$. 

The sums over the various $Y_{i}$'s are controled by lemma \ref{lemresexp} and
we are left with a sum over a tree that we perform much in the same way 
we did in
lemma \ref{lemresexp}. Indeed one can check that spatial decay appears through
factors of the form 
\begin{equation}
  \sum_{\Delta_{i}^{x}, \Delta_{i}^{y}, \Delta_{i}^{z}} 
    V_{\Delta_{l_{i}.y}} D_{\Lambda, \mu} \chi_{\Delta_{i}^{y}} R
    \chi_{\Delta_{i}^{z}} U_{\Lambda, \mu} \chi_{\Delta_{i}^{x}} D_{\Lambda,
    \mu} V_{\Delta_{l_{i}.x}}  
\end{equation}
thus we can extract decay in $d_{\Lambda}(l_{i}.y, l_{i}.x)$ time a bound in 
$\prod \|D_{\Lambda, \mu} V_{\Delta_{l_{i}.y}} D_{\Lambda, \mu}\|$ when
we combine all these factors. 

Yet we need some extra features to deal with the product of $O_{i}$'s,
each of them being bounded in norm by  $O(1) \, \mu^{-1} \|D_{\Lambda, \mu} 
V_{i} D_{\Lambda, \mu}\| \|D_{\Lambda, \mu} V_{k_{i}} D_{\Lambda, \mu}\|$ 
for some $k_{i}$. 

The factor $\mu^{-1}$ can be controled with a small fraction of the
probabilistic factor attached to the cube $\Delta_{i}$ 

If a given $D_{\Lambda, \mu} V_{i} D_{\Lambda, \mu}$ appears at a 
large power it has necessarily a large number of links
attached to it. Because of the tree structure, the links must go further and
further so that the decay of the links together with the gaussian measure 
allow to control the factorial coming from the accumulation of fields. 
This is quite standard and the reader can refer to \cite{Riv2} for instance. 

Finally we can write $T_{out, in}$ as a sum over polymers of the form 
\begin{eqnarray}
 T_{out, in} &=& \chi_{\Delta_{out}} (U_{\Lambda, \mu} + \delta T) 
   \chi_{\Delta_{in}} \\
  \delta T &=& \frac{\lambda^{c_{1}}}{[1 + 
    L^{-1} d_{\Lambda}(\Delta_{in}, \Delta_{out})]^{n_{3}}}
    \sum_{Y \in {\cal A}^{*}(\Delta_{in}, \Delta_{out})} \lambda^{c_{2}|Y|}
    \Gamma_{Y} T(Y) 
\end{eqnarray}
where $c_{1}$ and $c_{2}$ are small constant, $\Gamma_{Y}$ has decay in the
spatial extension of $Y$ and $\displaystyle 
\|\sum_{Y \in {\cal A}^{*}(\Delta_{in}, \Delta_{out})} T(Y)\|$ is bounded. 
\hfill 
$\blacksquare$ 


\section{Anderson model with an infra-red cut-off in dimension $d=2$} 
We are interested now in the particular case   
\begin{equation}
  H = -\Delta_{\eta} + \lambda \eta_{E} V \eta_{E} 
\end{equation} 
where
\begin{itemize}
  \item $\Delta_{\eta}^{-1}$ is a ultra-violet regularized inverse Laplacian, 
that we will note $-p^{2}$ 
  \item $\eta_{E}$ is an infra-red cut-off that forces 
$|p^{2} - E| \geqslant A \lambda^{2} |\log \lambda|^{2}$
  \item $V$ has covariance $\xi$ which is a ${\cal C}^{\infty}_{0}$
  approximation of a $\delta$-function 
  \item $M^{1/2}$ is an even integer greater than 2, and $j_{0} \in \NN$ is 
such that   
\begin{equation}
  M^{-j_{0}} \leqslant \inf_{\mbox{\scriptsize Supp}(\eta_{E})} |p^{2} - E| 
    \leqslant M^{-(j_{0}-1)}
\end{equation} 
\end{itemize}

For each $0\leqslant j \leqslant j_{0}$, we construct a smooth partition
of unity into cubes of side $M^{j}$ which form a 
lattice $I\!\! D_{j}$. It follows a decomposition of $V$ in fields 
$V_{\Delta_{j}}$ and we will assume
for simplicity that for $j<k$ and $\Delta_{k} \in I\!\! D_{k}$ 
\begin{equation}
  V_{\Delta_{k}} = \sum_{{\Delta_{j} \in I\!\! D_{j}} \atop { \Delta_{j} 
    \subset \Delta_{k}}} V_{\Delta_{j}}
\end{equation} 
even if it is not totally true because of irrelevant border effects. 

\subsection{The matrix model} 
We make a partition of unity according to the size of $p^{2}-E$ thanks to 
a function $\hat{\eta}$ which satisfies 
\begin{itemize}
  \item $\hat{\eta}$ is in ${\cal C}^{\infty}_{0}( \RR_{+})$ with 
    value in $[0, 1]$
  \item $\hat{\eta}$ has its support inside $[0, 2]$ and is equal to 
1 on $[0, 1]$
  \item the ${\cal L}^{\infty}$ norm of the derivatives of 
$\hat{\eta}$ does not grow too fast 
\end{itemize}
Then we construct
\begin{equation}
  \left\{
  \begin{array}{rcl}
    \hat{\eta}_{0}(p)&=&1-\hat{\eta} \left[M^{2} (p^{2}-E)^{2} 
      \right] \\
    \hat{\eta}_{j}(p)&=&\hat{\eta} \left[M^{2j}(p^{2}-E)^{2}\right] - 
      \hat{\eta} \left[M^{2(j+1)} (p^{2}-E)^{2} \right] 
      \quad \mbox{for } j > 0 \\
  \end{array}
  \right.
\end{equation}
In order to shorten expressions, we assume that 
\begin{equation}
  \eta_{E} = \sum_{j=0}^{j_{0}} \eta_{j}
\end{equation}

We expect that most of the physics will come from the neighborhood of the
singularity $p^{2}=E$ of the free propagator. 
As an operator in momentum space, $V$ has a
kernel $\hat{V}(p,q) \equiv \hat{V}(p-q)$. 
But since $p$ and $q$ have more or less the same norm, there are only two
configurations which give the sum $p-q$ (cf \cite{FMRT2}).  

\vspace{1em}
\centerline{\hbox{\psfig{figure=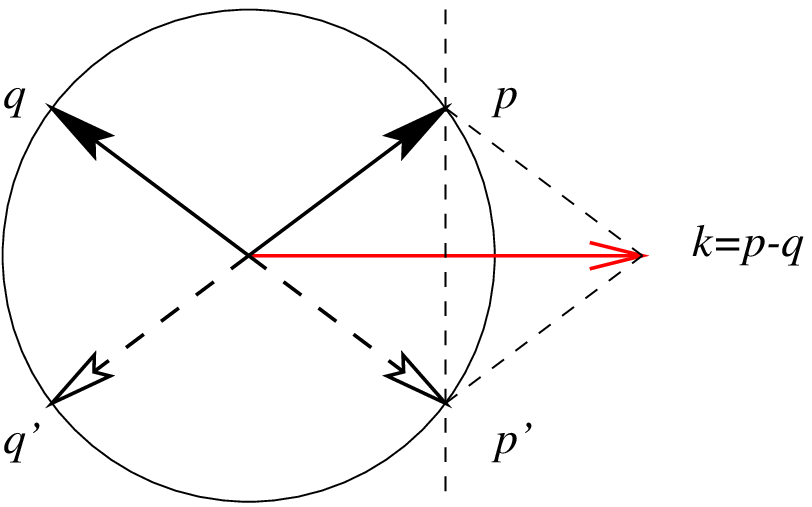,height=3cm}}}
\vspace{1em}

We can see this in another way. If we make perturbations and integrate on $V$
we will get Feynman graphs with four-legged vertices where the incoming
momenta have a fixed norm and must add to zero (or almost zero) because of
(approximate) translation invariance. Then the four momenta approximately 
form a rhombus which happens to be a parallelogram. 
It implies that they must be more or less opposite 2 by 2. Thus the problem 
looks like a vectorial model because the angular direction of the momentum is
preserved.   

In order to have this feature more explicit, we decompose the slice 
$\Sigma^{j} \equiv \mbox{Supp}(\hat{\eta}_{j})$ into $M^{j/2}$ angular
sectors. We introduce $\hat{\eta}_{S}$ with 
\begin{itemize}
  \item $\hat{\eta}_{S}$ is an even function in ${\cal C}^{\infty}_{0}(\RR)$ 
    with value in $[0, 1]$ 
  \item $\hat{\eta}_{S}$ has its support inside $[-1-\frac{1}{M},
    1+\frac{1}{M}]$ and is equal to 1 on $[-1,1]$ 
  \item $\hat{\eta}_{S} (1+x) = 1 - \hat{\eta} \left(1+\frac{1}{M}-x \right)$ 
    for $|x| \leqslant \frac{1}{M}$ 
  \item the ${\cal L}^{\infty}$ norm of the derivatives of $\hat{\eta}_{S}$ 
    does not grow too fast
\end{itemize}

Then we define $\theta_{j} = \pi M^{-j/2}$ and construct sectors
$S_{\alpha}^{j}$ of angular width $\theta_{j} (1+\frac{1}{M})$ centered around
$k_{\alpha} \equiv e^{i\alpha}$ (identifying $\RR^{2}$ and $\CC$), 
with $\alpha \in \theta_{j} \, \ZZ /_{\textstyle \! M^{\scriptstyle j/2} 
\ZZ}$. 
\begin{equation}
  \hat{\eta}_{j} = \sum_{\alpha} (\hat{\eta}_{\alpha}^{j})^{2} \quad 
    \mbox{where} \quad \left(\hat{\eta}_{\alpha}^{j}\right)^{2} 
    \left( |p| e^{i\theta}\right) 
    \equiv \hat{\eta}_{j}(|p|) \;
    \hat{\eta}_{S}\! \left(\frac{\theta-\alpha}{2 \theta_{j}}\right)
\end{equation} 

Afterwards, we define the operators $\eta_{\alpha}^{j}$'s by their kernel 
\begin{equation}
  \eta_{\alpha}^{j} (x, y) = \int \! dp \, e^{ip(x-y)}
    \hat{\eta}_{\alpha}^{j}(p)  
\end{equation}
They form a positive, self-adjoint partition of identity. 
\begin{equation}
  I = \sum_{j, \alpha} \left(\eta_{\alpha}^{j}\right)^{2} 
\end{equation}

We will map our problem to an operator-valued matrix problem with the
following lemma whose proof is quite obvious. 
\begin{lemma}
\ \\
Let ${\cal H}$ be an Hilbert space and suppose that we have a set of indices
${\cal I}$ and a partition of unity 
\begin{equation}
  I = \sum_{i \in {\cal I}} \eta_{i}^{2} 
\end{equation}
where $I$ is the identity in ${\cal L}({\cal H})$ and the $\eta_{i}$'s are
self-adjoint positive operators. 

For all $i \in {\cal I}$, we define 
\begin{equation}
  {\cal H}_{i} =\eta_{i} ({\cal H})
\end{equation}
then ${\cal H}$ and ${\cal L}({\cal H})$ are naturally isomorphic to
$\displaystyle \bigotimes_{i \in {\cal I}} {\cal H}_{i}$ and $\displaystyle 
{\cal L}\left(\bigotimes_{i \in {\cal I}} {\cal H}_{i} \right)$ thanks to 
\begin{eqnarray}
  x \in {\cal H} &\mapsto& (x_{i})_{i \in {\cal I}} \quad \mbox{ where } x_{i}
= \eta_{i} x \\ 
  A \in {\cal L}({\cal H}) &\mapsto& (A_{ij})_{i, j \in {\cal I}} \quad 
\mbox{ where } A_{ij} = \eta_{i} A \eta_{j}
\end{eqnarray}
\end{lemma}

In our case, we define ${\cal I}_{j}$ as the set of sectors in the slice
$j$ and ${\cal I} = \cup {\cal I}_{j}$ so that we can construct the
operator-valued matrices $\mbox{\bf V}_{\Delta}$'s as
\begin{equation}
  \left(\mbox{\bf V}_{\Delta}\right)_{\alpha \beta}^{jk} = 
    \eta_{\alpha}^{j} V_{\Delta} \eta_{\beta}^{k} 
\end{equation}

For a slice $\Sigma^{l}$, we define the enlarged slice 
\begin{equation}
  \bar{\Sigma}^{l} = \bigcup_{m \geqslant l} \Sigma^{m}
\end{equation}
Then an angular sector $S_{\alpha}^{l}$ of $\Sigma^{l}$ has a natural
extension into an angular sector $\bar{S}_{\alpha}^{l}$ of
$\bar{\Sigma}^{l}$ and we have the corresponding operator
$\bar{\eta}_{\alpha}^{l}$.


\subsection{Size of the V$^{jk}_{\Delta}$'s}
Let $\mbox{\bf V}_{\Delta}^{jk}$ be defined by 
\begin{equation}
  \left(\mbox{\bf V}_{\Delta}^{jk}\right)_{\alpha \beta}^{lm} = 
    \delta_{jl} \delta_{km} \eta_{\alpha}^{l} V_{\Delta} \eta_{\beta}^{m}  
\end{equation}
where the $\delta$ are Kronecker's ones. We can remark that 
\begin{equation}
  \left(V^{jk}_{\Delta}\right)^{\dag}=V^{kj}_{\Delta} \quad \Rightarrow \quad  
    \|V^{jk}_{\Delta}\| = \|V^{kj}_{\Delta}\| 
\end{equation}

Then we have the following large deviation result 
\begin{theorem} 
\label{thprobV} \  \\
There exists constants $C$ and $C_{\varepsilon}(\varepsilon)$ 
such that for all $\Lambda$, $j \leqslant k$, $a\geqslant 1$ and 
$\Delta \in \mbox{I}\! \mbox{D}_{k}$  
\begin{equation}
  I\!\!P_{\Lambda} \left(\| \mbox{\bf V}_{\Delta}^{jk}\| \geqslant 
    a C M^{-j/2}\right) \leqslant C_{\varepsilon} 
    e^{-(1-\varepsilon) a^{2} M^{(\frac{k}{2}- \frac{j}{3})}} 
\end{equation}
where $C_{\varepsilon}$ behaves like $1/\varepsilon$. 
\end{theorem}


\noindent {\it Proof} \\ 
We use the bound 
\begin{equation}
  \| \mbox{\bf V}^{jk}_{\Delta} \|^{2m_{0}} \leqslant
    \mbox{Tr} \left[\left(\mbox{\bf V}^{jk}_{\Delta}\right) 
    \left(\mbox{\bf V}^{jk}_{\Delta}\right)^{\dag} \right]^{m_{0}} 
\end{equation}
where 
\begin{eqnarray}
  \left(\mbox{\bf A}^{\dag}\right)_{\alpha \beta}^{lm} &=& 
    \left(\mbox{\bf A}_{\beta \alpha}^{ml} \right)^{\dag} \\
  \mbox{Tr \bf A} &=& \sum_{l, \alpha}  
    \mbox{tr \bf A}_{\alpha \alpha}^{ll}  
    = \sum_{l, \alpha} \int \!  
    \mbox{\bf A}_{\alpha \alpha}^{ll} (x,x) 
    \, dx
\end{eqnarray}
Thus for any $m_{0}$ 
\begin{eqnarray}
  I\!\!P \left(\| \mbox{\bf V}^{jk}_{\Delta} \| 
    \geqslant a C M^{-j/2} \right) &=&  
    \int \bbbone_{\left(\| \mbox{\bf V}^{jk}_{\Delta} \| \geqslant a
    C M^{-j/2} 
    \right)} d\mu_{\xi_{\Lambda}} (V) \\
  &\leqslant& \frac{1}{(a C M^{j/2})^{2m_{0}}} \int \| 
    \mbox{\bf V}^{jk}_{\Delta} \|^{2m_{0}} d\mu_{\xi_{\Lambda}} (V) \\ 
  &\leqslant& \frac{1}{(a C M^{-j/2})^{2m_{0}}} \! 
    \int \mbox{Tr} \left[\left(\mbox{\bf V}^{jk}_{\Delta}\right) 
    \left(\mbox{\bf V}^{jk}_{\Delta}\right)^{\dag} \right]^{m_{0}} 
    d\mu_{\xi_{\Lambda}} (V)
\end{eqnarray}

Let us note 
\begin{equation}
  {\cal I}_{m_{0}} \equiv 
  \mbox{Tr} \left[\left(\mbox{\bf V}^{jk}_{\Delta}\right) 
    \left(\mbox{\bf V}^{jk}_{\Delta}\right)^{\dag} \right]^{m_{0}} 
    \!\!\!\! = 
    \mbox{tr} \left[\sum_{\alpha} (\eta_{\alpha}^{j})^{2} \, 
    V_{\Delta} \sum_{\beta} (\eta_{\beta}^{k})^{2} \,  
    V_{\Delta} \right]^{m_{0}}
\end{equation}
We have the following lemma 
\begin{lemma}
\label{lemproba}
\ \\
There exists a constant $C$ such that for all $m_{0}$ we have the 
following bound
\begin{equation}
  < {\cal I}_{m_{0}}> \leqslant C^{2 m_{0}} M^{-jm_{0}} \left[1 +
M^{-m_{0}(\frac{k}{2}-\frac{j}{3})} m_{0}! \right] 
\end{equation}
\end{lemma}
This lemma is the core of the demonstration but its proof is quite long so
that we postpone it until the end of this part. It leads to 
\begin{equation}
    I\!\!P \left(\| \mbox{\bf V}^{jk}_{\Delta} \| \geqslant a C M^{-j/2} 
    \right) \leqslant a^{-2m_{0}} \left[1 + M^{-m_{0}(\frac{k}{2} -
    \frac{j}{3})} m_{0}!\right]  
\end{equation} 
We take $m_{0}= a^{2} M^{\frac{k}{2} - \frac{j}{3}}$ and use the rough bound
\begin{equation}
  n! \leqslant n^{(n+1)} e^{-(n-1)}
\end{equation}
to get the desired estimate. \hfill $\blacksquare$

In fact, in the proof of lemma \ref{lemproba} it is easy to see that 
$\eta_{j}$ and $\eta_{k}$ can be replaced by $\bar{\eta}_{j}$ and 
$\bar{\eta}_{k}$ with the same result.  
Furthermore, thanks to the locality of $V$ and to the decay of the $\eta$'s , 
the sum of several
$V_{\Delta}^{jk}$'s is more or less an orthogonal sum. More precisely, 
for any cube $\Delta_{0}$ we define $D_{m}(\Delta_{0})$ as the set of cubes
of $I\!\!D_{m}$ which are contained in $\Delta_{0}$. Then given two sets   
$\Omega_{1}$ and $\Omega_{2}$ and their smoothed characteristic functions
$\chi_{\Omega_{1}}$ and $\chi_{\Omega_{2}}$ we have 
\begin{lemma} 
\label{lemortho}
\ \\
For any $n$ and $C$ there is a constant $C_{n}$ such that 
for any $j \leqslant k$ and $\Delta_{0} \in I\!\!D_{k}$ 
\begin{eqnarray}
  \|\chi_{\Omega_{1}} \bar{\eta}_{j} V_{\Delta_{0}} \bar{\eta}_{k} 
    \chi_{\Omega_{2}} \|  
    &\leqslant&  
    \frac{C_{n}}{\left[1 + M^{-nj} d(\Omega_{1},\Delta_{0})^{n}
    \right]\left[1 + M^{-nk} d(\Omega_{2},\Delta_{0})^{n} \right]}  
    \nonumber \\ 
  && \quad \max \left[ \|\bar{\eta}_{j} 
    V_{\Delta_{0}} \bar{\eta}_{k}\|, 
    \sup_{{m<j} \atop {n<k}} \sup_{\Delta \in D_{m \wedge n}(\Delta_{0})} 
    \!\! M^{-C(j-m+k-n)} \|\eta_{m} V_{\Delta} \eta_{n}\| \right] 
\end{eqnarray}
where $m \wedge n = \min(m, n)$. 
\end{lemma}

\noindent {\it Proof} \\
We introduce $\chi_{\bar{\Delta}_{0}}$ a ${\cal C}^{\infty}_{0}$ function
equal to 1 on the support of $V_{\Delta_{0}}$ then we write 
\begin{equation}
  \chi_{\Omega_{1}} \bar{\eta}_{j} V_{\Delta_{0}} \bar{\eta}_{k} 
    \chi_{\Omega_{2}} = \chi_{\Omega_{1}} \bar{\eta}_{j}
    \chi_{\bar{\Delta}_{0}} \bar{\eta}_{j} V_{\Delta_{0}} \bar{\eta}_{k} 
    \chi_{\bar{\Delta}_{0}} \bar{\eta}_{k} \chi_{\Omega_{2}} 
    + \sum_{{m<j} \atop {n<k}} \chi_{\Omega_{1}} \bar{\eta}_{j}
    \chi_{\bar{\Delta}_{0}} \eta_{m} V_{\Delta_{0}} \eta_{n} 
    \chi_{\bar{\Delta}_{0}} \bar{\eta}_{k} \chi_{\Omega_{2}} 
\end{equation} 

Afterwards we introduce the sectors and the matrix formulation and we notice
that when we want to compute for instance the norm of the function 
$\eta^{n}_{\gamma} \chi_{\bar{\Delta}_{0}} \bar{\chi}^{k}_{\alpha}
\chi_{\Omega_{2}}$, momentum conservation tells us that we can convolve  
$\chi_{\bar{\Delta}_{0}}$ by a function which is restricted in momentum space
to the neighborhood of $S_{\gamma}^{n} - \bar{S}^{k}_{\alpha}$. In this way 
it is quite easy to see that we can extract at the same time spatial decay and
momentum conservation decay. \hfill $\blacksquare$


\subsection{Proof of theorem {\protect \ref{thprob2d}}}

Let $\Delta_{0} \in I\!\! D_{j_{0}}$, we call $X_{C_{x}, a}$ and 
$Y_{C_{y}, a}$ the events  
\begin{eqnarray}
  X_{C_{x}, a} &=& \left[ \exists j \leqslant k, \exists \Delta 
    \in D_{k}(\Delta_{0}) 
    \mbox{ s. t. } \| \bar{\eta}_{j} V_{\Delta} \bar{\eta}_{k} \| 
    \geqslant a C_{x} M^{-j/2} M^{\frac{j_{0}-k}{4}} \right] \\
  Y_{C_{y}, a} &=& \left[\| D_{\Lambda, \mu} \eta_{E} V_{\Delta_{0}} \eta_{E} 
    D_{\Lambda, \mu}\| \geqslant a C_{y} j_{0} M^{j_{0}/2} \right]
\end{eqnarray}
We will note $\bar{Z}$ the contrary event of $Z$.  

Theorem \ref{thprobV} tells us that 
\begin{eqnarray}
  I\!\! P(X_{C, a}) &\leqslant& \sum_{k} \sum_{j \leqslant k} 
    \sum_{\Delta \in D_{k}(\Delta_{0})} C' e^{-\frac{3}{4} a^{2}
    M^{\frac{j_{0} -k}{2}} M^{(\frac{k}{2} - \frac{j}{3})}} \\ 
  &\leqslant& \sum_{k} O(1) M^{2(j_{0}-k)} e^{-\frac{3}{4} a^{2}
    M^{\frac{j_{0}}{6}} M^{\frac{j_{0}-k}{3}}} \\
  &\leqslant& C_{1} e^{-\frac{1}{2} a^{2} M^{\frac{j_{0}}{6}}}
\end{eqnarray} 
One can see that thanks to lemma \ref{lemortho},  
$\bar{X}_{C, a}$ implies $\bar{Y}_{O(1) C, a}$. Thus if we call $C_{0} = O(1)
C$ 
\begin{equation}
  I\!\! P(Y_{C_{0}, a}) \leqslant I\!\! P(X_{C, a}) \leqslant 
    C_{1} e^{-\frac{1}{2} a^{2} M^{\frac{j_{0}}{6}}}
\end{equation}

Furthermore, if we work with respect to $\bar{X}_{C, a}$ which is stronger 
than $\bar{Y}_{C_{0}, a}$ everything goes as if one had 
\begin{equation} 
  \| \chi_{\Delta_{1}} D_{\Lambda, \mu} V_{\Delta_{2}} D_{\Lambda,
    \mu} \chi_{\Delta_{3}} \| \leqslant \frac{C_{n_{1}} \|D_{\Lambda, \mu}
    V_{\Delta_{2}} D_{\Lambda, \mu}\|}{\left[1 + L^{-1}
    d_{\Lambda}(\Delta_{1}, \Delta_{2})\right]^{n_{1}} \left[1 + L^{-1}
    d_{\Lambda}(\Delta_{2}, \Delta_{3})\right]^{n_{1}}}
\end{equation}
Thus we will be able to apply theorem \ref{thRE} with an effective coupling
constant 
\begin{equation}
  \lambda_{\mbox{\scriptsize eff}} = \lambda j_{0} M^{j_{0}/2}
\end{equation}
and a length scale $L = M^{j_{0}}$. 

If we want to make perturbations it is clever to perturb around the
expected Green's function without cut-off, {\it i.e.} we write 
\begin{equation}
  \frac{1}{p^{2}-E - i \mu +\lambda V} = \frac{1}{p^{2}-E - i \mu_{0} +\lambda
  V + i \delta \mu}
\end{equation}
where $\mu_{0}$ is the expected contribution of the tadpole given by the
self-consistent condition 
\begin{equation}
  \mu_{0} = \lambda^{2} \mbox{Im} \int \frac{1}{p^{2}-E-i \mu_{0}} dp
\end{equation}

Afterwards, when we compute the perturbative expansion, the tadpole with
cut-off will eat up a fraction $\lambda^{2} M^{j} \sim O(|\log \lambda|^{-2})$
of the counter-term so that 
\begin{equation}
  G \sim \frac{1}{p^{2}-E-i \eta_{E} O(\lambda^{2} |\log \lambda|^{-2})
  \eta_{E}}
\end{equation}
\hfill $\blacksquare$

In fact since the tadpole has a real part, it implies that we should
also renormalize the energy by a shift 
\begin{equation}
  \delta E = O \left( \lambda^{2} \log [\mbox{UV cut-off scale}] \right)
\end{equation}
 

\subsection{Proof of lemma {\protect \ref{lemproba}}}
We will note $J_{\alpha} \equiv (\eta_{\alpha}^{j})^{2}$,  
$K_{\beta} \equiv (\eta_{\beta}^{k})^{2}$ and $X$ as either $J$ or $K$. 

We can perform the integration on $V_{\Delta}$ so that $\left<{\cal I}_{m_{0}} 
\right>$ appears as a sum of Feynman graphs. 

\vspace{1em}
\begin{equation}
\left< {\cal I}_{m_{0}} \right> = \sum_{{\alpha_{1} 
\ldots \alpha_{m_{0}}} \atop {\beta_{1} \ldots \beta_{m_{0}}}} 
\raisebox{-2cm}{\psfig{figure=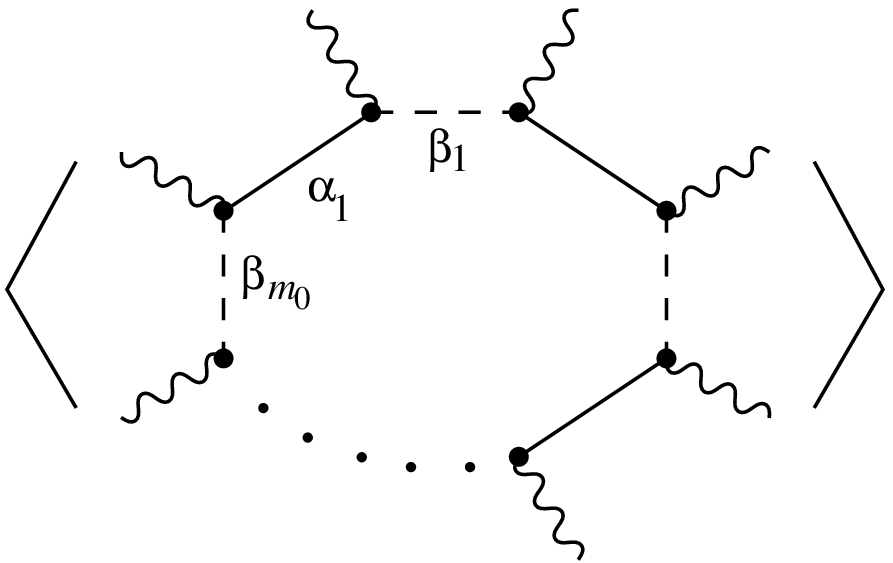,height=3cm}}
= \sum_{\cal G} {\cal A}({\cal G})
\end{equation}

\noindent where a solid line stands for a $J_{\alpha_{i}}$, 
a dashed line stands for a $K_{\beta_{i}}$ and a wavy line represents 
the insertion of a $V_{\Delta}$. 
In the following, we will prove the theorem in infinite volume with $V$
having a covariance $\delta$ in order to have shorter expressions. The proof
can then easily be extended to short range covariances and finite volume
except for the first few slices where one must pay attention to the
ultra-violet cut-off but this is irrelevant because it will cost only a factor
$O(1)$.

The integration on $V_{\Delta}$ consists in
contracting the wavy lines together, 
then both ends are identified and bear an extra $\chi_{\Delta}$ which 
restricts their position. 

The $X$'s will stand as propagators and the contraction of the
$V_{\Delta}$'s will give birth to 4-legged vertices. 


\subsubsection{Momentum conservation at vertices} 
First, we notice that if we note $\bar{\alpha}$ the opposite sector of
$\alpha$ 
\begin{equation}
  X_{\alpha}(x,y) = X_{\bar{\alpha}}(y,x)
\end{equation}

Then we put an orientation on each propagator, so that if a $X_{\alpha}$
goes from a vertex at $z$ to a vertex at $z'$ it gives a
$X_{\alpha}(z, z') = X_{\bar{\alpha}}(z', z)$, {\it i.e.} it is
equivalent to have an incoming $X_{\bar{\alpha}}$ at $z$ and an incoming 
$X_{\alpha}$ at $z'$. Now, for a given vertex with incoming propagators 
$X_{\alpha_{1}}$, $X_{\alpha_{2}}$, $X_{\alpha_{3}}$, $X_{\alpha_{4}}$, the
spatial integration over its position gives a term of the form 
\begin{equation}
  \Gamma_{\alpha_{1} \ldots \alpha_{4}} (x_{1}, x_{2}, x_{3}, x_{4}) = \int   
    \! X_{\alpha_{1}}(x_{1},z) X_{\alpha_{2}}(x_{2},z)
    X_{\alpha_{3}}(x_{3},z) X_{\alpha_{4}}(x_{4},z) 
    \chi_{\Delta}(z) \, dz
\end{equation} 

In momentum space, it becomes
\begin{equation}
  \Gamma_{\alpha_{1} \ldots \alpha_{4}} (p_{1}, \ldots, p_{4}) =
    X_{\alpha_{1}}(p_{1}) \ldots X_{\alpha_{4}}(p_{4}) 
    \int \! \chi_{\Delta}(k) \,  
    \delta(p_{1}+\ldots +p_{4}-k) \, dk
\end{equation}
where we use the same notation for a function and its Fourier transform. 

In $x$-space, $\chi_{\Delta}$ is a ${\cal C}^{\infty}_{0}$ function
with support inside a box of side $O(1) M^{j_{0}}$, it means that in
momentum space,  it is a ${\cal C}^{\infty}$ function with fast decay over a
scale $M^{-j_{0}}$. Thus for all $n$ there exists $C_{\chi}(n)$ such that 
\begin{equation}
  \left| \chi_{\Delta}(k)\right| \leqslant 
    \frac{C_{\chi}(n) M^{2j_{0}}}{\left(1+M^{2j_{0}} |k|^{2}\right)^{n+1}} 
\end{equation}

We make a decomposition of $\chi_{\Delta}$ 
\begin{equation}
  \chi_{\Delta}(k) = \sum_{s=0}^{j_{0}/2}
    \chi_{s}(k) 
\end{equation} 
where $\chi_{0}$ has its support inside the ball of radius 
$2M M^{-j_{0}}$, $\chi_{j_{0}/2} (\equiv \chi_{\infty})$ has its support 
outside the ball of radius $M^{j_{0}/2} M^{-j_{0}}$ and $\chi_{s}$ forces 
$|k|$ to be in the interval $[M^{s} M^{-j_{0}}; 2M M^{s} M^{-j_{0}}]$. 
In this way, we can decompose each vertex $v$ into a sum of vertices $v_{s}$,
where a vertex $v_{s}$ forces momentum conservation up 
to $O(1) M^{s} M^{-j_{0}}$ and has a factor coming from 
\begin{equation}
  \left| \chi_{s}(x) \right| \leqslant C'_{\chi}(n)M^{-sn} \times M^{-sn}
\end{equation}
We split the factor in order to have a small factor per vertex and yet retain
some decay to perform the sum on $s$.

\subsubsection{Tadpole elimination}

A graph will present tadpoles when two neighboring $V_{\Delta}$'s contract 
together thus yielding a $X(z,z)$. Suppose that
we have a $j$-tadpole then at the corresponding vertex we will have something
of the form
\begin{equation}
  \int \! \! dz \, \chi_{u}(z) \, K_{\beta} (x,z) \, J_{\alpha}(z,z) 
    \, K_{\beta'}(z,y) 
\end{equation}
Between the two $K$'s, momentum will be preserved up to 
$2M \, M^{u} M^{-j_{A}}$ which in most
case is much smaller than $M^{-j_{0}/2}$ so that $\beta'$ is very close to
$\beta$. Then we would like to forget about $J_{\alpha}$ by summing over
$\alpha$ and see the whole thing as a kind of new $K_{\beta}$. 
Now if per chance the new $K_{\beta}$ makes a tadpole we will erase it, 
and so on recursively. 

First, we define the propagators as propagators (or links) of order $0$ 
\begin{eqnarray}
  ^{0}\! J^{(0,0)}_{\alpha \alpha'} &=& \delta_{\alpha \alpha'}
    J_{\alpha} \\ 
  ^{0}\! K^{(0,0)}_{\beta \beta'}  &=& \delta_{\beta \beta'}
    K_{\beta} 
\end{eqnarray}

Then we define links of order $1$
\begin{eqnarray}
  ^{1}\! K^{(p,0)}_{\beta \beta'}(x,y) &=& \sum_{\beta_{1} \ldots 
    \beta_{p-1}} \sum_{{\alpha_{1} \ldots \alpha_{p}} \atop 
    {\alpha'_{1} \ldots \alpha'_{p}}} \int \! dz_{1} \ldots dz_{p} \, 
    K_{\beta}(x, z_{1}) \ ^{0}\! J^{(0,0)}_{\alpha_{1} \alpha'_{1}} 
    (z_{1}) \chi_{u_{1}}(z_{1}) \nonumber \\ 
    && \quad \quad  K_{\beta_{1}}(z_{1}, z_{2}) \ldots \ ^{0} \! 
    J_{\alpha_{p} \alpha'_{p}}^{(0,0)} 
    (z_{p}) \chi{u_{p}}(z_{p}) K_{\beta'}(z_{p}, y)
\end{eqnarray}
where we don't write the momentum conservation indices for shortness. We
have a similar definition  for $^{(1)}\! J^{(0,q)}_{\alpha \alpha'}$ 
(obtained by erasing $q$ $K$-tadpoles of order 0).
 
We will note $X^{(),t}$ to indicate that momentum is preserved up 
to $t M^{-j_{0}}$ between the leftmost and the rightmost $X$'s or
$X^{(),\infty}$ if momentum conservation is worse than $M^{-j_{0}/2}$.

Now, we can iterate the process in an obvious way. 
Yet, we must add an important restriction: we will erase a $X^{( ),\infty}$ 
tadpole only if it is attached to a $v_{\infty}$ vertex. 

\begin{lemma}
  \ \\
There exist constants $C_{1}$ and $C_{2}$ (independent of $j$ and $k$) 
such that for any tadpole obtained by erasing a total of $p$ $J$-tadpoles and
$q$ $K$-tadpoles we have the following bond
\begin{equation}
  \left|X_{\gamma \gamma'}^{(p,q)}(z, z)\right| \leqslant C_{1} C_{2}^{p+q} 
    M^{-pj-qk} M^{-3x/2} {\cal F}(X) \mbox{, } x = \left\{
    \begin{array}{l}
      \mbox{$j$ if $X=J$} \\
      \mbox{$k$ if $X=K$}
    \end{array}
    \right. 
\end{equation}
\end{lemma}
where ${\cal F}(X)$ is a small factor coming from the various
$\chi_{s}$ that appear in the expression of $X$. Thus, ${\cal F}$
gets smaller as momentum conservation gets worse.
 
\noindent {\it Proof} \\
First we will prove this result when momentum is well preserved, {\it i.e.} up
to $M^{-j_{0}/2}$ at worst, then we will see what has to be adapted when there
is a bad momentum conservation. 

The proof is by induction on the order of the tadpole. 
We define $C_{1}$, $C_{2}$ and $C_{3}$ such that 
\begin{eqnarray} 
  |X_{\gamma}(x, y)| &\leqslant& C_{1} M^{-3x/2} \\
  \sup_{x} \int \! dy \, \left| X_{\gamma}(x,y) \right| &\leqslant& C_{3} \\ 
  C_{2} &=& 9 C_{1} C_{3}
\end{eqnarray}
It is easy to see that for level 0 tadpoles 
\begin{equation}
  \left|^{0}\! X_{\gamma \gamma'} (z, z)\right| \leqslant C_{1} M^{-3x/2}  
    {\cal F}(X)
\end{equation}

Now, consider $^{m}\! J_{\alpha \alpha'}^{(p,q)}$ a $J$-tadpole of order 
$m$ and weight $(p, q)$ obtained by erasing $n$ $K$-tadpoles of order $m-1$ 
and weights $(p_{1}, q_{1})$, \ldots, $(p_{n}, q_{n})$. We have 
\begin{equation}
  p = p_{1}+ \ldots +p_{n} \quad \quad q = q_{1}+ \ldots + q_{n} + n
\end{equation} 
The expression of $^{m}\! J$ will be of the form 
\begin{eqnarray}
  ^{m}\! J_{\alpha \alpha'}^{(p,q)} (z, z) &=& \sum_{\alpha_{1} \ldots
    \alpha_{n-1}} \sum_{{\beta_{1} \ldots \beta_{n}} \atop 
    {\beta'_{1} \ldots \beta'_{n}}} \int \! dz_{1} \ldots dz_{n} \, 
    J_{\alpha}(z, z_{1}) \ ^{m-1}\! K_{\beta_{1} \beta'_{1}}^{(p_{1},
    q_{1})}(z_{1}, z_{1}) \nonumber \\ 
  && \hspace{-2em} \chi_{u_{1}}(z_{1}) \, J_{\alpha_{1}}(z_{1}, z_{2})
    \ldots K_{\beta_{n} \beta'_{n}}^{(p_{n}, q_{n})}(z_{n}, z_{n}) \, 
    \chi_{u_{n}}(z_{n}) J_{\alpha'}(z_{n}, z)
\end{eqnarray} 

Since we supposed that we have momentum conservation up to $M^{-j_{0}/2}$, the
$\alpha_{i}$'s will be either $\alpha_{i-1}$ or one of its neighbors and 
$\beta'_{i}$ will be either $\beta_{i}$ or one of its neighbors. Thus the
sum on sector attribution will give a factor $3^{2n-1} M^{nk/2} \leqslant
9^{n} M^{nk/2}$. 

We have $n+1$ $J$'s but only $n$ spatial integrations because we have a
tadpole. This gives a factor $C_{3}^{n} C_{1} M^{-3j/2}$ (we
forget about the momentum conservation factor for the moment). 

Finally the $^{m-1}\! K$'s bring their factor so that 
\begin{eqnarray}
  \left|^{m}\! J_{\alpha \alpha'}^{(p,q)} (z, z) \right| &\leqslant &
    9^{n} M^{nk/2} C_{3}^{n} C_{1} M^{-3j/2} \nonumber \\
  && \left(9C_{1} C_{3}\right)^{\sum p_{i}+ \sum q_{i}} 
    M^{-j \sum p_{i} -k \sum q_{i}}
    \left(C_{1} M^{-3k/2}\right)^{n} {\cal F}(X) \nonumber \\
  &\leqslant& C_{1} \left(9 C_{1} C_{3}\right)^{\sum p_{i}+ \sum q_{i} + n} 
     M^{-j \sum p_{i} -k (\sum q_{i} +n)} M^{-3j/2}  {\cal F}(X)
\end{eqnarray}
which is precisely what we want. Then we can do the same for the $^{m}\! K$'s.
 
Now we must consider the cases with bad momentum conservation. 
First, let us suppose that momentum conservation is bad overall for 
$^{m}\! J$ but was good for
the  $^{m-1}\! K$'s, then the previous argument will work except if there
are some $v_{\infty}$ vertices. In this case we will have to
pay a factor $M^{j/2}$ to find the following $\alpha_{i}$ instead of a factor
3. But from the corresponding $\chi_{\infty}$ we have a small factor 
\begin{equation}
  \frac{1}{1+ M^{j_{0}N'/2}}
\end{equation}
from which we can take a fraction to pay the $M^{j/2}$ and retain a small
factor for ${\cal F}(X)$. 

Finally, if a $^{m-1}\! K$ has a bad momentum conservation it is
necessarily attached to a $v_{\infty}$ vertex (otherwise we
would not erase it). In this case we must pay a factor $M^{j/2} M^{k/2}$ (to
find $\beta'_{i}$ and $\alpha_{i}$) but again we can take a fraction of the
factor of $\chi_{\infty}$ to do so.

When tadpole elimination has been completed, we have erased $t_{j}$ 
$J$-tadpoles and $t_{k}$ $K$-tadpoles and we are left
with $m_{0}' = m_{0} - t_{j} - t_{k}$ vertices linked together by 
$m'_{0}$ $J$'s and $m'_{0}$ $K$'s (a tadpole which has not been erased 
being seen as a propagator). 

For a  $X_{\alpha \alpha'}^{(p, q)}(x,y)$, it is quite easy to see that to
integrate on $y$ with fixed $x$ amounts more or less to the same problem for 
$O(1)^{p+q} X_{\alpha}$ and that to find $\alpha'$ knowing $\alpha$ costs a
factor $O(1)^{p+q}$.   

		
\subsubsection{Sector conservation at the vertex}
\begin{lemma}
\label{sconserv}
\ \\
\it Let ($\bar{S}_{\alpha_{1}}^{l}$, \ldots, $\bar{S}_{\alpha_{4}}^{l}$) be a
quadruplet of sectors of the enlarged slice $\bar{\Sigma}^{l}$ and 
$0 \leqslant r \leqslant O(1)M ^{l/2}$ such that 
there are $p_{1} \in \bar{S}^{l}_{\alpha_{1}}$, \ldots, $p_{4} \in
\bar{S}^{l}_{\alpha_{4}}$ verifying
\begin{equation}
  |p_{1}+ \ldots + p_{4}| \leqslant r M^{-l} \quad \mbox{with} \quad
\end{equation}  

Then we can find $\{\alpha, \alpha', \beta, \beta'\} = 
\{\alpha_{1}, \ldots, \alpha_{4}\}$ satisfying
\begin{equation}
\left\{
\begin{array}{lcr}
|\alpha' - \bar{\alpha}| &\leqslant& (a \sqrt{r}+ b) \, M^{-l/2} \\  
|\beta' - \bar{\beta}| &\leqslant& (a \sqrt{r} + b) \, M^{-l/2} 
\end{array}
\right.
\end{equation}
where $a$ and $b$ are some constants independent of $l$ and $r$. 
\end{lemma}


\noindent {\it Proof} \\
If we can prove the result for $l \geqslant O(1)$ then we will be able to
enlarge the result to any $l$ provided maybe we take some slightly bigger $a$
and $b$. Therefore we assume that this is the case in the following.
  
We define $(\alpha, \alpha', \beta, \beta')$ by 
\begin{itemize}
\item $\{\alpha, \alpha', \beta, \beta'\} = \{ \alpha_{1}, \ldots, 
  \alpha_{4} \}$ 
\item $\alpha = \alpha_{1}$
\item $\displaystyle |\alpha -\beta| = \min_{i \in \{2, 3, 4\}} 
  |\alpha -\alpha_{i}|$
\item $|\bar{\alpha} - \alpha'| \leqslant |\bar{\alpha} - \beta'|$
\end{itemize}
Then, if $|\alpha -\beta| \leqslant |\alpha' - \beta'|$ we exchange 
$(\alpha, \beta)$ and $(\alpha', \beta')$.

A sector $\bar{S}^{l}_{\gamma}$ is included in a tube, of center $k_{\gamma} \equiv
e^{i\gamma}$ and whose direction is orthogonal to the direction $\gamma$, of 
size
\begin{equation}
\left\{
\begin{array}{lcl}
   \mbox{length}&:& L= \pi M^{-l/2} (1+{\textstyle \frac{2}{M}}) \\
   \mbox{width}&:&2 M^{-l}
\end{array}
\right.
\end{equation}

We define
\begin{eqnarray}
   k_{\alpha \beta} &=& k_{\alpha} + k_{\beta} = 2  \cos \left(
     \frac{\alpha-\beta}{2}\right) e^{i\frac{\alpha+\beta}{2}} \equiv
     2 \cos x \, e^{i\theta} \equiv r e^{i\theta} \\
   k_{\bar{\alpha}' \bar{\beta}'} &=& - k_{\alpha' \beta'} \equiv 2 \cos
     \bar{x}' \, e^{i\bar{\theta}'}
\end{eqnarray}

If we can prove that 
\begin{equation}
\left\{
\begin{array}{lcr}
|\bar{x}' - x| &\leqslant& (a' \sqrt{r}+ b') \, M^{-l/2} \\  
|\bar{\theta}' - \theta| &\leqslant& (a' \sqrt{r} + b') \, M^{-l/2} 
\end{array}
\right.
\end{equation}
then we will be able to conclude, with $a=2a'$ and $b=2b'$.

It is easy to check that by construction, we have 
\begin{itemize}
  \item $0 \leqslant \bar{x}' \leqslant x$
  \item $|\alpha - \beta| \leqslant \frac{2 \pi}{3} \Rightarrow \cos x
    \geqslant \frac{1}{2}$ 
\end{itemize}

We have a trivial bound 
\begin{equation}
  |k_{\alpha \beta} - k_{\bar{\alpha}' \bar{\beta}'}| \leqslant  2 \tan 
  \left|\frac{\bar{\theta}' - \theta}{2}\right| \leqslant r M^{-l} + 2L
    + 4 M^{-l} \equiv R
\end{equation}

Therefore 
\begin{equation}
  |\bar{\theta}' - \theta| \leqslant 2 \tan \left|\frac{\bar{\theta}' - 
    \theta}{2}\right| \leqslant R \leqslant O(1) M^{-l/2} 
\end{equation}

We can see that $\theta$ is very well conserved. 

If $\sin x \leqslant (a_{1} \sqrt{r} + b_{1}) M^{-l/2}$ then 
$|x- \bar{x}'| \leqslant x \leqslant (a_{2} \sqrt{r} + b_{2}) M^{-l/2}$.

Otherwise, let us remark that $\bar{S}^{l}_{\alpha} + \bar{S}^{l}_{\beta}$ is 
at a distance at most $2 M^{-l}$ from a rhombus $R_{\alpha \beta}$ of center
$k_{\alpha \beta}$ and of diagonals 
\begin{equation}
\left\{
\begin{array}{l}
  2 L \sin x \quad \mbox{in the direction {\bf u}}_{r} \equiv 
    \frac{\alpha + \beta}{2} \\
  2 L \cos x \quad \mbox{in the direction {\bf u}}_{\theta} \equiv 
    \frac{\alpha + \beta}{2} + \frac{\pi}{2}
\end{array}
\right.
\end{equation}

Then, $R_{\alpha \beta} - R_{\bar{\alpha}' \bar{\beta}'}$ is at a distance at
most $4 M^{-l}$ from a rectangle ${\cal R}$ of center $k_{\alpha \beta} -
k_{\bar{\alpha}' \bar{\beta}'}$ and of sides
\begin{equation}
\left\{
\begin{array}{l}
  L_{r} = 2L \left( \sin x + \sin \bar{x}' \cos |\bar{\theta}' - \theta| + 
    \cos \bar{x}' \sin |\bar{\theta}' - \theta| \right)  
    \mbox{ in {\bf u}}_{r} \\
  L_{\theta} = 2L \left( \cos x +\cos \bar{x}' \cos |\bar{\theta}' - \theta| 
    + \sin \bar{x}' \sin |\bar{\theta}' -\theta| \right) 
    \mbox{ in {\bf u}}_{\theta} 
\end{array}
\right.
\end{equation}

Since $|\bar{\theta}' - \theta| \leqslant O(1) M^{-l/2}$, we have $|\cos
(\bar{\theta}' - \theta) -1| \leqslant O(1) M^{-l}$. 
We define a $z$ axis in the direction $\mbox{\bf u}_{r}$. 
$(k_{\bar{\alpha}' \bar{\beta}'} -
k_{\alpha \beta})$ has a $z$ coordinate $2 (\cos \bar{x}' - \cos x) +
O(1) M^{-l}$. This leads to the condition
\begin{eqnarray}
  2 |\cos \bar{x}' - \cos x| &\leqslant& r M^{-l } + 
    2L (\sin x + \sin \bar{x}') +b_{3} M^{-l} \\ 
    &\leqslant& (r + b_{3}) M^{-l} + 4 L \sin x 
\end{eqnarray}

Let us note that $|\cos (x-u) - \cos x|$ is an increasing function of $u$ 
and that we have 
\begin{equation}
  \cos (x-u) - \cos x = \sin x \, u - \frac{1}{2} \cos x \, u^{2} +
    u^{3} \varepsilon(u) \quad \mbox{with } |\varepsilon(u)| \leqslant
    \frac{1}{6}  
\end{equation}
We take $u = (\sqrt{r} +b_{4}) M^{-l/2} + 2L \equiv (\sqrt{r} + b'_{4}) 
M^{-l/2} \leqslant O(M^{-l/4})$. 
\begin{eqnarray}
  |\cos (x - u) - \cos x| &\geqslant&  
    \sin x \, u - \left(\frac{1}{2} + \frac{u}{6} \right) u^{2} \\
  &\geqslant& 2L \sin x + \sin x (\sqrt{r} +b_{4}) M^{-l/2} 
    - (\sqrt{r} +b'_{4})^{2} M^{-l} 
\end{eqnarray}
Since we are in the case $\sin x \geqslant (a_{1} \sqrt{r} +b_{1})
M^{-l/2}$, for $a_{1}$ and $b_{1}$ large enough we will have 
\begin{equation}
  2 |\cos (x-u) - \cos x| \geqslant (r + b_{3}) M^{-l} + 4L \sin x 
\end{equation}

Therefore we must have $|\bar{x}' -x| \leqslant u \leqslant (\sqrt{r}+b'_{4})
M^{-l/2}$ which allows us to conclude.
\hfill 
$\blacksquare$


\subsubsection{Size of a graph}

The previous section shows that, at each vertex, momenta come approximately
by pairs of opposite sectors. Thus, for all the vertices which haven't been
erased by the tadpole elimination process, we can choose by a factor $3$ how
to pair the sectors. Then we split the vertices in two half-vertices according
to this pairing. We represent graphically this as

\vspace{1em}
\centerline{\hbox{\psfig{figure=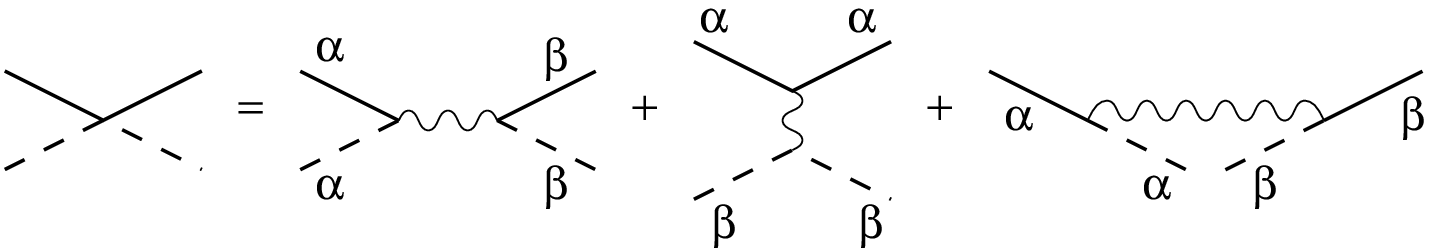,width=10cm}}}
This gives $3^{m'_{0}}$ (split) graphs that we will consider as our basic
graphs in the following.  

A graph is decomposed into a number of {\em momentum cycles} connected
together by wavy lines. We will follow those cycles to fix momentum sectors. 
Finding the enlarged sectors (of level $j$) will cost a factor $M^{j/2}$ per
cycle times a constant per vertex. Then we will pay an extra $M^{(k-j)/2}$ for
each $K$ propagator to find its sector. 

We define $c$ the total number of momentum cycles that we decompose into $t$
tadpoles, $b$ bubbles (with 2 vertices) and $l$ large cycles (with 3 or more
vertices). We have 
\begin{eqnarray}
  t + b + l &=& c \\
  \label{momcycle} t + 2b + 3l &\leqslant& 2 m'_{0}
\end{eqnarray}
and the sector attribution costs 
\begin{equation}
  \label{Cpainting}
 {\cal A}_{1} = C^{m_{0}} M^{cj/2} M^{m'_{0} (k-j)/2}
\end{equation} 
Notice that the constant has an exponent $m_{0}$ because of the tadpole
elimination process. 

The spatial integration of the vertices will be made with the short $J$ links
whenever possible. We can decompose each graph into $J$-cycles linked 
together by $K$ links (because there are $2$ incoming $J$'s at each vertex),
this allows to integrate all the vertices but one per cycle with a $J$
link. The total cost is (noticing that the last vertex is integrated in the
whole cube $\Delta$) 
\begin{equation}
\label{Cintegrate}
  {\cal A}_{2} = C^{m_{0}} M^{3(m'_{0} - c')j/2} 
M^{3 (c'-1) k/2} M^{2k} 
\end{equation}
where $c'$ is the total number of short cycles that we decompose
into $t'$ short tadpoles, $b'$ short bubbles and $l'$ large short cycles.  
\begin{eqnarray}
  t' + b' + l' &=& c' \\
  \label{shortcycle}t'+ 2b' + 3 l' &\leqslant& m'_{0}
\end{eqnarray} 

The scaling of the tadpoles and the propagators give a factor 
\begin{equation}
\label{Cscaling}
  {\cal A}_{3} = M^{-jt_{j} - k t_{k}} M^{-3 m_{0}' (j+k)/2} 
\end{equation}

Tadpoles that have been obtained by erasing a few vertices (say
$O(M^{j_{0}/4})$ for instance) will have an extra small factor because they
strongly violate momentum conservation, we can take it to be a power of
$M^{-j_{0}/4}$. Tadpoles with higher weights will not
have this good factor but we will see that they bring a better
combinatoric. The $t$ momentum tadpoles will consist in $t_{1}$ low weight
ones and $t_{2}$ others while the $t'$ short tadpoles split into $t'_{1}$ low
weight ones and $t'_{2}$ others. We can manage to have a factor 
\begin{equation}
  {\cal A}_{4} = M^{-2 t_{1} j_{0}} M^{-2 t'_{1} j_{0}} 
\end{equation} 

If we have a short bubble we will have four incoming long propagators whose
momenta must add up to zero up to $x M^{-j_{0}}$. If we apply lemma
\ref{sconserv}, we can see that knowing 3 of these momenta it cost only a
factor $O(1) \sqrt{x}$ to find the fourth momentum sparing us a factor
$M^{(k-j)/2}$ obtained by na{\"\i}vely fixing first the enlarged sector at slice
$j$. If the bubble has a weight $p$, {\it i.e.} the two short propagators have
been obtained after erasing $p$ vertices, and a momentum conservation worse
than $O(1) p M^{-j_{0}}$ then the small factor of bad momentum conservation
will pay for the $O(1) \sqrt{p}$. We will have $b'_{1}$ such good bubbles, each
of them bringing a factor $M^{-(k-j)/2}$. 
In addition, we will have $b'_{3}$ bad bubble of
weight greater than $M^{(k-j)}$ for which we earn nothing and $b'_{2}$ bad
bubbles of weight $p_{i}$ bringing a factor $C \sqrt{p_{i}} M^{-(k-j)/2}$. 
This gives a factor 
\begin{equation}
  {\cal A}_{5} = M^{-(b'_{1}+b'_{3})(k-j)/2} C^{b'_{3}} \prod_{i} \sqrt{p_{i}}
\end{equation}

Finally we have the following bound for the contribution of a graph
\begin{eqnarray}
  |{\cal A}({\cal G})| &\leqslant& {\cal A}_{1} \ldots {\cal A}_{5} \nonumber
    \\  
  & \leqslant & C^{m_{0}} M^{cj/2} M^{m'_{0}(k-j)/2}
    M^{3(m'_{0}-c')j/2} M^{3(c'-1)k/2} M^{2k} 
    M^{-jt_{j} - kt_{k}} \nonumber \\ 
  && M^{-3m'_{0}(j+k)/2} M^{-2 j_{0}(t_{1} +t'_{1})} 
    M^{-(b'_{1}+b'_{2})(k-j)/2} \prod_{i} \sqrt{p_{i}} \\ 
  &\leqslant& C^{m_{0}} M^{k/2} M^{cj/2} M^{3c'(k-j)/2}
    M^{-m'_{0}(k+j/2)} M^{-jt_{j} -kt_{k})} M^{-2j_{0}(t_{1}+t_{1}')} 
    \nonumber \\ 
  && M^{-(b'_{1}+b'_{2})(k-j)/2} \prod_{i} \sqrt{p_{i}} 
\end{eqnarray}

If we use equations (\ref{momcycle}) and (\ref{shortcycle}) we obtain 
\begin{eqnarray}
  |{\cal A}({\cal G})| &\leqslant& C^{m_{0}} M^{k/2} 
    M^{-m'_{0}(k+j)/2} M^{-j t_{j} -k t_{k}}  
    M^{-m'_{0}j/6}  M^{bj/6} \nonumber \\ 
\label{graphbond}  && M^{t_{2}j/3} M^{t'_{2} (k-j)} 
    M^{b'_{3}(k-j)/2} M^{-t_{1}(2j_{0}-j/3)} M^{-t'_{1}(2j_{0}+j-k)} 
    \prod_{i} \sqrt{p_{i}} 
\end{eqnarray}

we will take $m_{0} \geqslant M^{k/6}$ so that $M^{k/2} \leqslant
 C^{m_{0}}$. Furthermore, $t$ and $t'$ are at most
equal to $M^{-j_{0}/4} m_{0}$ thus $M^{kt} \leqslant C^{m_{0}}$. It allows us
 to rewrite the bound  
\begin{equation}
  |{\cal A}({\cal G})| \leqslant C^{m_{0}} M^{-m'_{0}(k+j)/2} M^{-j t_{j} -k
    t_{k}} M^{-(m'_{0}-b) j/6} M^{b'_{3}(k-j)/2} \prod_{i} \sqrt{p_{i}}
\end{equation}


\subsection{Graph counting}
\begin{lemma}
\ \\
Let $T(p)$ be the number of ways to contract $2p$ adjacent $V$'s so as to
make only generalized tadpoles. We have 
\begin{equation}
\label{Ntadpole} T(p) = \frac{(2p) !}{p! (p+1)!}  
\end{equation}
\end{lemma}

\noindent {\it Proof} \\
It is easy to see that a good contraction scheme, {\it i.e} one that gives
only generalized tadpoles, corresponds to have no crossing contractions. It
means that if we label the fields $V_{0} \ldots V_{2p-1}$ according to their 
order and if $V_{i}$ and $V_{j}$ contract respectively to $V_{k}$ and $V_{l}$
then 
\begin{equation}
  i < j \Rightarrow k<j \mbox{ or } k>l
\end{equation} 
We have $T(1)=1$. For $p>1$, we contract first $V_{0}$ to some $V_{i}$. 
$V_{1}, \ldots, V_{i-1}$ will necessarily contract among themselves making 
only generalized tadpoles and so will do $V_{i+1}, \ldots, V_{2p-1}$. 
Thus $i$ is necessarily odd and we have 
\begin{equation}
\label{Trecursion} T(p) = \sum_{k=0}^{p-1} T(k) T(p-1-k) 
\end{equation} 
where by convention $T(0)=1$. 

We introduce the generating function
\begin{equation}
  t(z) = \sum_{p = 0}^{\infty} T(p) z^{p}
\end{equation}
The recursion formula (\ref{Trecursion}) can be translated into an equation
for $t$ which is 
\begin{equation}
 t(z) = z t^{2}(z) + 1 
\end{equation} 
whose resolution yields 
\begin{equation}
  t(z) = \frac{1 + \sqrt{1-4z}}{2z} \mbox{ or } \frac{1 - \sqrt{1-4z}}{2z} 
\end{equation}
Since the second solution is analytic around $z=0$, we can take it as $t(z)$
and the coefficients of its power expansion will give us $T(p)$. An easy
computation leads to the desired formula. 
\hfill $\blacksquare$

\begin{lemma}
\ \\
The number ${\cal N}_{M}(B)$ of graphs with $B$ possible momentum bubbles 
obtained in the contraction of a cycle of $2M$ $V$'s has the following bound
\begin{equation}
  {\cal N}_{M}(B) \leqslant C^{M} (M-B)!
\end{equation}
\end{lemma}

\noindent {\it Proof} \\ 
First lets us remark that bubbles come in chains (possibly with a tadpole at
one end) of two possible types
\begin{itemize}
  \item type 1: \raisebox{-0.6em}{\psfig{figure=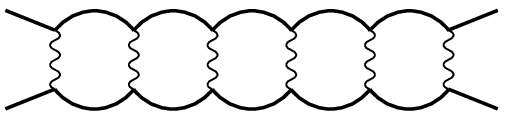,height=0.75cm}} 
  \item type 2: \raisebox{-0.6em}{\psfig{figure=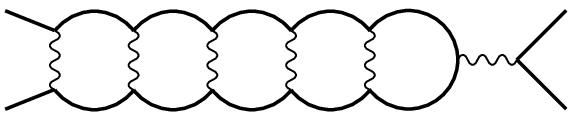,height=0.75cm}} 
\end{itemize}
where a solid line stands here either for a $J$ or a $K$. 

We have two special cases 
\begin{itemize}
  \item \raisebox{-0.6em}{\psfig{figure=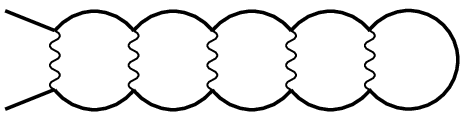,height=0.75cm}} which can
    be seen as a type 1 chain 
  \item \raisebox{-0.6em}{\psfig{figure=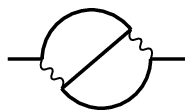,height=0.75cm}} which 
    can generate only one momentum bubble so that we can see it as a
    type 2 chain of length 1.  
\end{itemize}

Having chosen the $V$'s there are only two contraction schemes that yield a
type 1 chain and a unique contraction scheme for a type 2 chain. If
we fix explicitly the subgraphs corresponding to $B$ bubbles 
and contract the remaining $V$'s in any way we will get all the desired graphs
plus some extra ones so that we can bound ${\cal N}_{M}(B)$. 

We construct $r_{1}$ type 1 chains of lengths $\beta_{1}$, \ldots,
$\beta_{r_{1}}$ and $r_{2}$ type 2 chains of lengths $\gamma_{1}$, \ldots,
$\gamma_{r_{2}}$. We set 
\begin{eqnarray}
  B_{1} &=& \sum_{i} \beta_{i} \\ 
  B_{2} &=& \sum_{i} \gamma_{i} \\ 
  B &=& B_{1} + B_{2}
\end{eqnarray}

To count the contraction schemes, first we cut the cycle of $2M$ $V$'s into a
sequence of $V$'s (there are $2M$ ways to do so). Then it is easy to check
that in order to build a type 1 chain we must choose two sets ${\cal B}_{i}$
and $\bar{\cal B}_{i}$ of $\beta_{i}+1$ adjacent $V$'s while for a type 2 we 
need a set ${\cal D}_{i}$ of $2 \gamma_{i}+1$ adjacent
$V$'s. We distribute those $2 r_{1}+r_{2}$ objects in   
$(2M-2B_{1}-2r_{1}-2B_{2}-r_{2})+2r_{1}+r_{2}$ boxes in an ordered way, 
and for the $\mbox{i}^{th}$ type 1 chain the
respective order of ${\cal B}_{i}$ and $\bar{\cal B}_{i}$ will fix the
contraction scheme. Then, there remain $2M -2B-2r_{1}$ $V$'s to contract so
that we have the following number of configurations 
\begin{equation}
  {\cal N}_{M}(B) \leqslant 2M \!\! \sum_{B_{1}+B_{2}=B} \sum_{{r_{1} 
    \leqslant B_{1}} 
    \atop {r_{2}\leqslant B_{2}}} \sum_{{\beta_{1}+ \ldots \beta_{r_{1}}=B_{1},
    \beta_{i} \geqslant 1} 
    \atop {\gamma_{1}+ \ldots \gamma_{r_{2}}=B_{2}, \gamma_{i} 
    \geqslant 1}} \frac{1}{r_{1}!} 
    \frac{1}{r_{2}!} \frac{(2M-2B)!}{(2M-2B-2r_{1}-r_{2})!} 
    (2M - 2B -2r_{1}-1)!! 
\end{equation} 
We can compute this
\begin{eqnarray}
  {\cal N}_{M}(B) &\leqslant& 2M \!\! \sum_{B_{1}+B_{2}=B} 
    \sum_{{r_{1}\leqslant
    B_{1}} \atop {r_{2}\leqslant B_{2}}}
    {B_{1}-1 \choose r_{1}-1} {B_{2}-1 \choose r_{2}-1}
    \frac{[2(M-B)]!}{[2(M-B)-2r_{1}-r_{2}]! (2r_{1})! r_{2}!} \nonumber \\
  &&\frac{(2 r_{1})!}{r_{1}!} \frac{[2(M-B-r_{1})]!}{2^{M-B-r_{1}}
    (M-B-r_{1})!} \\ 
  &\leqslant& 2M \!\! \sum_{B_{1}+B_{2}=B} \sum_{{r_{1}\leqslant
    B_{1}} \atop {r_{2}\leqslant B_{2}}} 
    {B_{1}-1 \choose r_{1}-1} {B_{2}-1 \choose r_{2}-1} 3^{2(M-B)} 
    2^{2r_{1}} r_{1}! 2^{M-B-r_{1}} (M-B-r_{1})! \\
 &\leqslant& 2M \!\! \sum_{B_{1}+B_{2}=B} 2^{B_{1}-1} 2^{B_{2}-1} 9^{M-B} 
   2^{M} (M-B)! \\
 &\leqslant& 2M (B+1) 18^{M} (M-B)!  
\end{eqnarray}
\hfill $\blacksquare$


\subsection{Bounds}
Now we can achieve the proof of lemma \ref{lemproba} in bounding 
\begin{equation}
  < {\cal I}_{m_{0}}> \leqslant \sum_{\cal G} |{\cal A}({\cal G})| 
\end{equation}

In order to compute this sum, we fix first $t_{j}$, $t_{k}$
and $\bar{b}$, where $\bar{b}$ is the number of possible momentum bubbles and
therefore is greater than $b$. 

Then we define the set $\Omega(t_{j}, t_{k}, \bar{b}, n, q_{1}, \ldots,
q_{n})$ has the set of graphs with the corresponding $t_{j}$, $t_{k}$ and
$\bar{b}$ and for which the erased tadpoles form $n$ sets of $2 q_{i}$
adjacent $V$'s. We can write 
\begin{equation}
  < {\cal I}_{m_{0}}> \leqslant \sum_{t_{j}, t_{k, \bar{b}}} 
    \sum_{n=1}^{t_{j}+t_{k}} \frac{1}{n!} 
    \sum_{{q_{1}, \ldots q_{n}} \atop {q_{i} \geqslant 1}} 
    \sum_{{\cal G} \in \Omega(t_{j}, t_{k}, \bar{b}, n, q_{1}, \ldots q_{n})} 
    \prod (q_{i}+1) \prod \frac{1}{q_{i}+1} |{\cal A}({\cal G})| 
\end{equation}
To bound $|{\cal A}({\cal G})|/ \prod(q_{i}+1)$ we notice that when a graph
has a bad bubble of weight $p_{i}$ it means that we have erased two set of
generalized tadpoles $q_{1_{i}}$ and $q_{2_{i}}$ on the two propagators of the
bubble with $q_{1_{i}} + q_{2_{i}}=p_{i}$. Thus we have a corresponding
factor $(q_{1_{i}}+1)^{-1} (q_{2_{i}}+1)^{-1}$ which control the bad factor
$\sqrt{p_{i}}$ of the bad bubble so that
\begin{equation}
  < {\cal I}_{m_{0}}> \leqslant \sum_{t_{j}, t_{k, \bar{b}}} 
    \sum_{n=1}^{t_{j}+t_{k}} \frac{1}{n!} 
    \sum_{{q_{1}, \ldots q_{n}} \atop {q_{i} \geqslant 1}} 
    \sum_{{\cal G} \in \Omega(\ldots)} 
    \prod (q_{i}+1) C^{m_{0}} M^{-m'_{0}(k+j)/2} M^{-jt_{j}-kt_{k}}
    M^{-(m'_{0}-\bar{b})j/6} 
\end{equation}

The number of graphs in $\Omega(\ldots)$ has the following bound 
\begin{equation}
  {\cal N}\left[\Omega(\ldots)\right] \leqslant 2^{m_{0}} \prod T(q_{i})
3^{m'_{0}} C^{m'_{0}} (m'_{0}-\bar{b})! \leqslant C^{m_{0}} \prod
\frac{2^{2q_{i}}}{q_{i}+1} (m'_{0}-\bar{b})!
\end{equation}

This leads to 
\begin{eqnarray}
 < {\cal I}_{m_{0}}> &\leqslant& \sum_{t_{j}, t_{k, \bar{b}}} 
    C^{m_{0}} M^{-m'_{0}(k+j)/2} M^{-jt_{j}-kt_{k}}
    M^{-(m'_{0}-\bar{b})j/6} (m'_{0}-\bar{b})! \nonumber \\ 
  && \quad \quad \sum_{n=1}^{t_{j}+t_{k}}  
     \frac{1}{n!} \sum_{{q_{1}+ \ldots q_{n} = t_{j}+t_{k}} 
    \atop {q_{i} \geqslant 1}} 1 \\  
  &\leqslant&  \sum_{t_{j}, t_{k, \bar{b}}} 
    C^{m_{0}} M^{-m'_{0}(k+j)/2} M^{-jt_{j}-kt_{k}}
    M^{-(m'_{0}-\bar{b})j/6} (m'_{0}-\bar{b})! \sum_{n=1}^{t_{j}+t_{k}} 
    {t_{j}+t_{k}-1 \choose n-1} \\ 
  &\leqslant&  \sum_{t_{j}, t_{k, \bar{b}}} 
    C^{m_{0}} M^{-m'_{0}\frac{(k+j)}{2}} M^{-jt_{j}-kt_{k}}
    M^{-(m'_{0}-\bar{b}) \frac{j}{6}} (m'_{0}-\bar{b})!
\end{eqnarray}

Summing on $t_{j}$ and $t_{k}$ is equivalent to sum over $m'_{0}$ and $t_{k}$
with $t_{j}=m_{0}-t_{k}-m'_{0}$. The sum over $\bar{b}$ is roughly evaluated
by taking the supremum over $\bar{b}$, the result depends whether $m'_{0}$ is
greater than $M^{j/6}$ or not. 
\begin{eqnarray}
  < {\cal I}_{m_{0}}> &\leqslant& \sum_{m'_{0}\leqslant M^{j/6}}  
    \sum_{t_{k}}  
    C^{m_{0}} M^{-m'_{0}(k+j)/2} M^{-jt_{j}-kt_{k}} \nonumber \\
  && \quad + \sum_{m'_{0}> M^{j/6}} \sum_{t_{k}}  
    C^{m_{0}} M^{-m'_{0}(k+j)/2} M^{-jt_{j}-kt_{k}} 
    \max \left[1, M^{-m'_{0}j/6} (m'_{0}-\bar{b})! \right]  \\ 
 &\leqslant& C^{m_{0}} M^{-m_{0}j} \sum_{m'_{0}, t_{k}} 
   M^{-m'_{0}(k-j)/2} M^{-t_{k}(k-j)} \left[ 1 + 1_{(m'_{0}> M^{j/6})} 
   M^{-m'_{0}j/6} (m'_{0}-\bar{b})! \right]
\end{eqnarray}
 
Finally, the sum over $t_{k}$ is easy and we bound the sum over $m'_{0}$ 
by finding the supremum. One can check that it gives
the announced result. \hfill $\blacksquare$


\section{Conclusion}
Understanding the effect of perturbations on the free spectrum of
Hamiltonian operators is an outstanding challenge. We think that for the
two-dimensional case, this paper can help to control the model up to
quite close to the scale of the expected ``mass'' ({\it i.e.} imaginary part) 
so that in studying the full model, one can focus on the thin slice of 
momentum $p^{2}-E \sim \lambda^{2}$. Then we think that a key of the problem 
lies in the fact that in
this case the potential is very close (in momentum space) to a large hermitian 
random matrix with almost independent entries and therefore we can connect our
problem to the much better understood domain of random matrices or
equivalently use the vector model picture of the problem. 

One can note that in dimension $d=3$, the interpretation in term of random
matrices can also be helpful (cf. \cite{MPR}), but then one has to deal with
constrained matrices whose entries are no longer independent.


\section{Acknowledgments}
This paper is part of a larger program in collaboration with J. Magnen
and V. Rivasseau to apply rigorous renormalization group methods to the study
of disordered systems, the ultimate goal being to devise tools to
construct extended states in weakly disordered systems in dimension
$d \geqslant 3$. They played an important part in getting these
results and I thank them very much for their help.



\end{document}